\magnification=1200
%
\baselineskip=18pt
\newcount\sec
\sec=0
\newcount\dummy

%
%
%
\newcount\num
\def\eqnum{\global\advance\num by 1 \eqno(\the\sec.\the\num )}
\def\eqnlbl#1{\eqnum\xdef #1{(\the\sec.\the\num )}}
\def\eqnumapp#1{\global\advance\num by 1 \eqno({\rm #1}.\the\num )}
\def\eqnumapplbl #1#2 {\eqnumapp#1\xdef #2{({\rm #1}.\the\num )}}
\def\eqalnum{\global\advance\num by1 (\the\sec .\the\num)}
\def\eqalnumlbl#1{\eqalnum\xdef #1{(\the\sec.\the\num )}}

%
\newcount \refno
\refno=0
\xdef\gobble#1{}
\def\ifundefined #1
  {\expandafter\ifx\csname\expandafter\gobble\string#1\endcsname\relax}

\def\ref #1{$
   \ifundefined {#1}
   \global\advance\refno by 1
   \the\refno
   \xdef #1 {\the\refno}
   \else #1 \fi   $}


\def\writeref{
 \def\ref ##1{$
   \ifundefined {##1}
      \global\advance\refno by 1
      \xdef ##1 {\the\refno\string ##1}
      ##1
   \else ##1 \fi   $}     }

\font\Bigrm=cmr10 scaled\magstep2
\font\bigrm=cmr10 scaled\magstep1




\def\heading#1{\goodbreak\bigskip\line{\bf #1\hfil}
        \nobreak\smallskip\nobreak\noindent}

\def\smallheading#1{\goodbreak\bigskip\line{\it #1\hfil}
        \nobreak\noindent}
\def\noskipsmallheading#1{\line{\it #1\hfil}\nobreak\noindent}

\def\leaderfill{\leaders\hbox to 1em{\hss.\hss}\hfill}

\def\roothalf{{1\over \sqrt 2}}
\def\half{{\scriptstyle{1\over 2}}}




\def\Z{\hbox{Z$\!\!$Z}}

\def\GeV{\hbox{GeV}}
\def\TeV{\hbox{TeV}}

\null
\line{\hfil SHEP-92/93-21 \quad}
\line{\hfil hep-ph/9308309 \quad}
\vfill
\line{\hfil {\Bigrm Radiative Corrections to Higgs Boson Masses in the}\hfil}
\line{\hfil {\Bigrm Next-to-Minimal Supersymmetric Standard Model} \hfil}

\vskip 40pt
\line{\hfil {\bigrm T. Elliott, S.F. King and P.L. White} \hfil}
\vskip 10pt
\line{\hfil {\it Physics Department, University of Southampton,}\hfil}
\line{\hfil {\it Southampton SO9 5NH, UK.}\hfil}
\vskip 40pt

\heading{\hfil Abstract }
We perform a systematic study of radiative corrections to the masses
of the Higgs bosons in the minimal supersymmetric standard model
(MSSM) augmented by a single gauge singlet, the so-called
next-to-minimal supersymmetric standard model (NMSSM). Our method is
based on the one-loop effective potential and includes effects of top
quark, squark, Higgs and Higgsino loops. We discuss the
renormalisation group flows of Yukawa couplings and the upper bound on
the lightest CP-even neutral Higgs boson mass as a function of the
heavier stop mass and top mass. We then give a general discussion of
Higgs boson phenomenology including radiative corrections.  We survey
as much of the parameter space of the Higgs sector of the NMSSM as is
practicable, and analyse the full spectrum of Higgs masses and
couplings in these regions of parameter space.  Characteristic
signatures of the NMSSM such as light charged bosons and weakly
coupled neutral scalars are discussed, as are the relative sizes of
the various radiative corrections. The MSSM is also discussed as a
limiting case of the NMSSM for comparison.

\vfill
\eject

\num=0
\sec=1

\heading{1 Introduction}
The most widely studied extension to the standard model is
supersymmetry (SUSY) [\ref{\reviews}], in which the particles of the
theory are supplemented by the inclusion of their superpartners. SUSY
requires the introduction of a non-minimal Higgs sector with at least
two doublets [\ref{\hhguide}].  An important question is how heavy the
lightest neutral CP-even supersymmetric Higgs boson, $h$, can be
within the framework of supersymmetric grand unified theories (SUSY
GUTs). In SUSY GUTs all the Yukawa couplings are constrained to remain
perturbative in the region $M_{SUSY}\sim 1$ TeV to $M_{GUT}\sim
10^{16}$ GeV. This constraint provides a maximum value at low energies
for those Yukawa couplings which are not asymptotically-free, and is
obtained from the renormalisation group (RG) equations together with
the boundary conditions that the couplings become non-perturbative at
$M_{GUT}$ -- the so-called ``triviality limit''. In the minimal
supersymmetric standard model (MSSM) [\ref{\reviews}, \ref{\hhguide}],
the triviality limits provide a useful bound on the top quark mass
$m_t$. The upper bound on the $h$ mass, $m_h$, in the MSSM, including
radiative corrections, has recently been the subject of much
discussion [\ref{\okada}, \ref{\erz}, \ref{\brignole}, \ref{\eahd},
\ref{\haber}].

However the MSSM is not the most general low energy manifestation of
SUSY GUTs. It is possible that SUSY GUTs give rise to a low energy
theory which contains an additional gauge singlet field, the so called
next-to-minimal supersymmetric standard model (NMSSM) [\ref{\fayet},
\ref{\NMSSM}, \ref{\dl}].  In all supersymmetric models, including
those with non-minimal Higgs sectors, there is an upper bound on the
lightest CP-even scalar Higgs mass, which is generally larger for
non-minimal models [\ref{\eq}, \ref{\kkw}]. Although the NMSSM involves a
single gauge singlet, the estimates of
the bound are applicable to a model with an arbitrary number of Higgs
singlets, since the singlet fields may always be redefined so that
only one of them couples to the doublets and only this field
contributes to the upper two-by-two mass matrix which gives the bound.
This argument cannot be applied to models which contain extra
non-singlet degrees of freedom (although these usually give a lower
value for the bound) or to the description of the whole spectrum in
any model other than the NMSSM.

There has recently been much interest in radiative corrections to
Higgs boson masses in the NMSSM [\ref{\eltraub}, \ref{\ellind},
\ref{\el}, \ref{\terV}, \ref{\eqtwo}, \ref{\ekwone}, \ref{\ekwtwo},
\ref{\pandita}]. The two approaches that have been considered are the
renormalisation group (RG) approach [\ref{\terV}, \ref{\eqtwo},
\ref{\ekwone}],  and the one-loop effective potential approach
[\ref{\eltraub}, \ref{\ellind}, \ref{\el}, \ref{\ekwtwo},
\ref{\pandita}]. In the RG approach one derives an effective low energy
Higgs potential at a low energy scale $\mu$ from SUSY matching
conditions at a scale $M_{SUSY}$.  The squarks and other sparticles
are usually assumed to be degenerate at $M_{SUSY}$, and the
logarithmic radiative corrections are efficiently summed by the RG
equations for the case of either one [\ref{\terV}, \ref{\eqtwo}] or
two [\ref{\ekwone}] light Higgs doublets.  This approach, although
convenient, fails to pick up non-logarithmic corrections, and in
general becomes complicated when a general squark spectrum is
considered [\ref{\haber}]. The one-loop effective potential
[\ref{\cw}, \ref{\sher}] has been used in the MSSM to estimate the
radiative corrections due to a general squark spectrum [\ref{\erz}]
and these calculations have been repeated in the NMSSM [\ref{\ellind},
\ref{\el},
\ref{\ekwtwo}]. However in the NMSSM there are other corrections due to
Higgs loops which may become large due to the presence of large
couplings and trilinear soft parameters, and it is one of the purposes
of the present paper to consider such contributions to the one-loop
effective potential.

In this paper we shall perform a systematic study of radiative
corrections to Higgs boson masses in the NMSSM, using the one-loop
effective potential. We shall consider the effects of all particles
which couple through relatively large (order one) couplings in the
model including the effects of loops of top quarks, stop squarks,
Higgs bosons, and Higgsinos. The top and stop corrections have been
calculated before [\ref{\ellind}, \ref{\el}, \ref{\ekwtwo}] and we
include them in our analysis for completeness.  The Higgs and Higgsino
contributions to the effective potential have not been calculated
before, and are dealt with here by numerical techniques since their
contribution has no simple analytic form.

Having developed the techniques for dealing with radiative corrections
in the NMSSM we then apply these techniques in two different ways. The
first application is to the problem of the bound on the lightest
CP-even Higgs boson mass, which was discussed above.  More generally
we give a phenomenological discussion of Higgs boson masses and
couplings over as much of the parameter space of the NMSSM as it is
feasible to consider.  Our goal is to understand the behaviour of the
Higgs boson spectrum as each of the parameters in turn is varied, and
to give some examples of phenomenological signatures which would
enable the NMSSM to be distinguished from the MSSM.  It is important
to stress that we shall not impose GUT scale constraints on the soft
SUSY-breaking parameters, so that our analysis is of a general low
energy phenomenological nature.

We shall also present an RG analysis of the dimensionless Yukawa
couplings of the model which will turn out to provide a useful guide
to the typical values that these parameters may take at low energy.

The plan of the rest of this paper is as follows. Section 2 introduces
the NMSSM and the tree-level potential and Higgs boson mass matrices.
Section 3 presents an RG analysis of the dimensionless couplings
between $M_{GUT}$ and $M_{SUSY}$. In section 4 we discuss in general
terms the radiative corrections to the Higgs mass matrix using the
one-loop effective potential. In section 5 we use numerical methods to
obtain an upper bound on the lightest CP-even neutral Higgs boson mass
as a function of the various parameters. This is compared to the case
in the minimal model.  In section 6 we discuss Higgs boson masses and
couplings in some detail and discuss the effect of radiative corrections.
Section 7 concludes the paper.

\num=0
\sec=2
\heading{2 The Next-to-Minimal Supersymmetric Standard Model }
The most commonly considered supersymmetric model, the MSSM, has, in
addition to the usual
matter and gauge particle content, a Higgs sector containing two Higgs
doublet superfields $H_1$ and $H_2$. The superpotential is then of the
form
$$
W_{MSSM}=h_uQu^cH_2 + h_dQd^cH_1 + h_eLe^cH_1 - \mu H_1H_2 ,
\eqnlbl{\MSSMsuppotl} $$
where generation indices are understood, $H_1H_2=H_1^0H_2^0-H_1^-H_2^+$,
with $H_1^T=(H_1^0, H_1^-)$, $H_2^T=(H_2^+, H_2^0)$,
and the rest of the notation is conventional. In this model the physical
Higgs spectrum consists of two CP-even and one CP-odd neutral scalars,
and the lightest neutral scalar $h$ has a mass which is bounded at tree
level by $m_h^2\le m_Z^2$ [\ref{\hhguide}].

In the NMSSM [\ref{\fayet},\ref{\NMSSM},\ref{\dl}]
the particle content of the MSSM is supplemented by a gauge
singlet superfield, $N$. The superpotential is given by
$$
W_{NMSSM}=h_uQu^cH_2 + h_dQd^cH_1 + h_eLe^cH_1 + \lambda NH_1H_2 -
    {k\over 3}N^3.
\eqnlbl{\suppotl} $$
The cubic term in $N$ is necessary to avoid a Peccei-Quinn symmetry
which would force the existence of a light pseudo-Goldstone mode once
the symmetry is broken. However there still remains a $\Z_3$ symmetry
under which all the matter and Higgs superfields $\Phi$ transform as
$\Phi\to\alpha\Phi$ where $\alpha^3=1$ [\ref{\NMSSM}]. Note that we have
eliminated $\mu$. This can be justified on the grounds of naturalness,
and its inclusion would only complicate our analysis. The  gauge singlet
field acquires a vacuum expectation value (vev) which plays the role of
the mass parameter $\mu$ in the MSSM.

Unlike the case in non-supersymmetric models, radiative corrections do
not generate large masses of order the cut-off scale (which for SUSY
GUTS is essentially the unification scale $10^{16}$GeV), although the
inclusion of singlets may cause the destabilisation of the hierarchy if
there are strong couplings to super-heavy particles such as Higgs colour
triplets [\ref{\stabone}]. This is however strongly dependent on the
structure of the model at the GUT scale, and so we shall not discuss it
here. Recently it has also been noted that the inclusion of
non-renormalisable operators suppressed by powers of the Planck mass may
lead to the introduction of non-logarithmic divergences, which in turn
can generate large mass terms for the Higgs bosons of the electroweak
theory [\ref{\stabtwo}].  This effect can only occur in the case of a
model which has a singlet, and requires that gravity violate the $\Z_3$
symmetry which is respected by the renormalisable operators of the
theory.  The coefficients of such operators have not yet been calculated
and their size and importance is unclear.

In our analysis we shall drop all quark and lepton Yukawa couplings
apart from that of the top quark so that the superpotential reduces to
$$
W_{NMSSM} \approx h_t Q H_2 t^c + \lambda N H_1 H_2 - {k\over 3} N^3 ,
\eqnum
$$
where the superfield $Q^T=(t_L,b_L)$ contains the left--handed top and
bottom quarks, and $t^c$ contains the charge conjugate of the
right--handed top quark.
Adopting the usual convention of using the same symbols for both
component Higgs fields and superfields, the fields
$H_1$, $H_2$, and $N$ develop vevs
which may be assumed to be of the form
$$
<H_1> = \left( \matrix{ \nu_1 \cr 0 \cr}  \right), \ \ \
<H_2> = \left( \matrix{ 0 \cr \nu_2 \cr} \right), \ \ \
<N> = x =r\nu ,
\eqnum
$$
where $\nu_1$, $\nu_2$ and $x$ are real, $\sqrt{\nu_1^2+\nu_2^2}
= \nu =174$ GeV, and $\tan\beta=\nu_2/\nu_1$.
The low energy physical spectrum of the Higgs scalars consists of three
CP-even neutral states, two CP-odd neutral states, and two charged
scalars. A third CP-odd state is a Goldstone mode which becomes the
longitudinal component of the $Z^0$, while a further two charged
degrees of freedom become those of the $W^\pm$s.

In addition to the potential which can be derived from the
superpotential in the usual manner, there is a soft supersymmetry
breaking potential of the form
$$\eqalign{
V_{\rm soft} =
   & h_tA_t\tilde Q\tilde t^cH_2 - \lambda A_{\lambda}NH_1H_2
         -  {k\over 3}A_kN^3 \cr
   &\ + m_{H_1}^2 \vert H_1 \vert^2 + m_{H_2}^2 \vert H_2 \vert^2
              + m_{N}^2 \vert N \vert^2 \cr
   &\ + m_{T}^2 \vert\tilde t^c \vert^2 + m_{B}^2 \vert\tilde b^c \vert^2
              + m_{Q}^2 \vert\tilde Q \vert^2 \cr
}\eqnlbl{\Vsoft}
$$
which leads to the full low energy Higgs potential $V_0$, where
$$\eqalign{
V_0  = \,
     & \half {\lambda}_1({H_1}^{\dagger}H_1)^2
      + \half {\lambda}_2({H_2}^{\dagger}H_2)^2
      + ({\lambda}_3 +{\lambda}_4)
              ({H_1}^{\dagger}H_1)({H_2}^{\dagger}H_2) \cr
     & - {\lambda}_4|{H_2}^{\dagger}H_1|^2
      + {\lambda}_5|N|^2|H_1|^2 + {\lambda}_6|N|^2|H_2|^2 \cr
     & + {\lambda}_7({N^{\ast}}^2H_1H_2 + N^2H_1^{\ast}H_2^{\ast})
                   + {\lambda}_8|N|^4 \cr
     & + {m_1}^2|H_1|^2 + {m_2}^2|H_2|^2 + {m_3}^2|N|^2 \cr
     & -  m_4(H_1H_2N + H.c.)- {1\over 3} m_5(N^3 + H.c.). \cr
}\eqnlbl{\potl}
$$

The quartic couplings ${\lambda}_i$ and the mass parameters $m_i$ must
satisfy the following boundary conditions at $M_{SUSY}$
$$\eqalign{
\lambda_1 = \lambda_2= {(g_2^2 + g_1^2)\over 4},&\qquad
\lambda_3= {(g_2^2 - g_1^2)\over 4} \cr
\lambda_4= \lambda^2- {g_2^2 \over 2},&\qquad
\lambda_5=\lambda_6={\lambda}^2, \cr
\lambda_7=-{\lambda}k,& \qquad {\lambda}_8=k^2, \cr
m_1^2=m_{H_1}^2, \qquad m_2^2=&m_{H_2}^2, \qquad m_3^2=m_{N}^2, \cr
m_4 = {\lambda}A_{\lambda},& \qquad m_5=kA_k \cr
}\eqnlbl{\bndcond} $$
where $g_1$, $g_2$ are the usual $U(1)$ and $SU(2)$ gauge couplings of
the standard model.

Since we have three minimisation conditions,
${\partial V_0\over \partial \nu_i}=0$ and
${\partial V_0\over\partial x}=0$,
we may eliminate three of the unknown parameters of the theory, which we
choose to be $m_1$, $m_2$, and $m_3$, in favour of the three vevs.  The
remaining masses $m_4$ and $m_5$ are related to the parameters
$A_{\lambda}$ and $A_k$ at $M_{SUSY}$ as above. Because we know $\nu$
(from the $W$ mass) the Higgs sector of this model is now parametrised
in terms of the six parameters $\lambda$, $k$, $\tan\beta$, $r$, $m_4$,
and $m_5$. We shall take the parameters $\lambda$, $k$, $A_{\lambda}$
and $A_k$ to be real, and $\lambda$, $k$ to be positive, which is a
sufficient condition for the vacuum to conserve CP and leads to a choice
of vacuum in which all the vevs $x,\nu_1,\nu_2$ are real and positive
[\ref{\NMSSM}].

The range of the parameters is restricted by the condition that the
vacuum does not break QED in the Higgs sector, which is not automatic
in the NMSSM, and is equivalent to the condition that $m_c^2\geq 0$,
where $m_c$ is the mass of the physical charged Higgs $H^{\pm}$.
Another similar problem is that the vacuum may break QCD in the squark
sector, but this does not occur for sufficiently small $A_t$
[\ref{\NMSSM}, \ref{\ds}]. Slepton vevs will not be discussed here
since they can be avoided by an appropriate choice of soft parameters.

The full $10\times 10$ mass squared
matrix ${\cal M}^2$ for the scalar fields is
simple to derive by expressing all of the fields in terms of their real
scalar and pseudo-scalar parts
$$\eqalign{
H_1^0=\,& \roothalf (\varphi_1 + i \varphi_4) \cr
H_2^0=\,& \roothalf (\varphi_2 + i \varphi_5) \cr
N    =\,& \roothalf (\varphi_3 + i \varphi_6) \cr
H_1^-=\,& \roothalf (\varphi_7 - i \varphi_9) \cr
H_2^+=\,& \roothalf (\varphi_8 + i \varphi_{10}) \cr
}\eqnlbl{\phidefs} $$
and then using
$$
{\cal M}^2_{ij}= {\partial^2 V_0 \over \partial\varphi_i\partial\varphi_j}
\eqnlbl{\fullMsq}
$$
When evaluated at the vevs,
$$\eqalign{
<\varphi_1>=&\sqrt{2}\nu_1 \cr
<\varphi_2>=&\sqrt{2}\nu_2 \cr
<\varphi_3>=&\sqrt{2}x \cr
<\varphi_i>=&0 \quad\forall i\neq 1,2,3 \cr
}\eqnum $$
${\cal M}^2$ breaks down to consist
entirely of zeros except in one $3\times 3$ block for the CP-even, one
$3\times 3$ for the CP-odd, and two $2\times 2$ blocks for the charged
mass matrices (the CP-odd matrix and each of the charged matrices have
one zero eigenvalue corresponding to a Goldstone mode, and the two
non-zero charged eigenvalues are equal). Thus the tree-level neutral
CP-even (scalar) mass squared symmetric matrix,
in the basis $\{H_1,H_2,N\}$, is
$$
M^2  = \left( \matrix{
  2\lambda_1 \nu_1^2 &  2(\lambda_3+\lambda_4) \nu_1 \nu_2
        & 2\lambda_5 x \nu_1\cr
  2(\lambda_3+\lambda_4) \nu_1 \nu_2 &  2\lambda_2 \nu_2^2
        & 2\lambda_6 x \nu_2 \cr
  2\lambda_5 x \nu_1 & 2\lambda_6 x \nu_2
        & 4\lambda_8 x^2 - m_5x \cr
} \right)
$$
$$
\qquad \qquad +
\left( \matrix{
\tan \beta [m_4x-\lambda_7x^2] & - [m_4x-\lambda_7x^2] &
     -{{\nu_2}\over {x}} [m_4x-2\lambda_7x^2] \cr
- [m_4x-\lambda_7x^2] & \cot \beta [m_4x-\lambda_7x^2] &
     -{{\nu_1}\over {x}} [m_4x-2\lambda_7x^2] \cr
-{{\nu_2}\over {x}} [m_4x-2\lambda_7x^2] & -{ \nu_1\over x}
  [m_4x-2\lambda_7x^2] & {{\nu_1 \nu_2}\over {x^2}} [m_4x] \cr}
 \right)
\eqnlbl{\scalmatrix}
$$
Similarly the tree-level neutral CP-odd (pseudoscalar) mass-squared
symmetric matrix, in the basis $\{H_1,H_2,N\}$, is
$$
\tilde{M}^2 =
\left( \matrix{
\tan \beta [m_4x-\lambda_7x^2] & [m_4x-\lambda_7x^2] &
       {{\nu_2}\over {x}} [m_4x+2\lambda_7x^2] \cr
[m_4x-\lambda_7x^2] & \cot \beta [m_4x-\lambda_7x^2] &
       {{\nu_1}\over {x}} [m_4x+2\lambda_7x^2] \cr
{{\nu_2}\over {x}} [m_4x+2\lambda_7x^2] &
{{\nu_1}\over {x}} [m_4x+2\lambda_7x^2] &
   3m_5x + {{\nu_1 \nu_2}\over {x^2}} [m_4x-4\lambda_7x^2]\cr}
\right).
\eqnlbl{\pseudmatrix}
$$
Finally the tree-level charged mass-squared matrix,
in the basis $\{H_1,H_2\}$, is
$$
M_c^2= \left( \matrix{
                      \tan\beta &     1       \cr
                          1     & \cot\beta   \cr
                                                 } \right)
  (m_4x-\lambda_7x^2-\lambda_4\nu_1\nu_2).
\eqnlbl{\chargedmatrix}
$$

In the limit $\lambda ,k \rightarrow 0$,
$x \rightarrow \infty$ with $\lambda x$ and $kx$ held fixed,
the $N$ components do not mix with the $H_1,H_2$ components
in the mass matrices in \scalmatrix   and \pseudmatrix .
This is just the MSSM limit of the NMSSM.

\num=0
\sec=3
\heading{ 3 Renormalisation Group Analysis }
Now let us consider which values may be taken by the
dimensionless couplings $\lambda$, $k$ and $h_t$.
Above some scale at which supersymmetry becomes a good symmetry, and
defining $ t=\log\mu$, where $\mu$ is the renormalisation scale, the RG
equations for these couplings [\ref{\ds}] are given by
$$\eqalign{
8 \pi^2 {\partial\lambda\over\partial t} = \, &
        ( 2\lambda^2+k^2+{3\over 2}h_t^2
             - {3\over 2} g_2^2-{1\over 2}g_1^2)\lambda \cr
8 \pi^2 {\partial k\over\partial t} = \, &
        ( 3\lambda^2+3k^2)k \cr
8 \pi^2 {\partial h_t\over\partial t} = \, &
        ( {1\over 2}\lambda^2+3h_t^2 - {8\over 3}g_3^2
             - {3\over 2} g_2^2-{13\over 18}g_1^2)h_t \cr
}\eqnum$$
where $g_3$ is the QCD coupling. Following the analysis of reference
[\ref{\bs}], they may be written in the suggestive form
$$\eqalign{
8 \pi^2 {\partial\hfill\over\partial t}
 \bigl ({\lambda^2\over h_t^2} \bigl ) = \, &
        ( 3\lambda^2+2k^2-3h_t^2 +{16\over 3}g_3^2
             + {4\over 9}g_1^2)  \bigl ({\lambda^2\over h_t^2} \bigl ) \cr
8 \pi^2 {\partial\hfill\over\partial t}
 \bigl ({k^2\over h_t^2} \bigl ) = \, &
        ( 5\lambda^2+6k^2-6h_t^2 +{16\over 3}g_3^2 +{3\over 2}g_2^2
             + {13\over 9}g_1^2)  \bigl ({k^2\over h_t^2} \bigl ) \cr
}\eqnlbl{\RGratios} $$
These have three fixed points in the gaugeless ($g_i=0,\ \forall i$) limit:
$$\eqalign{
 \bigl ({\lambda^2\over h_t^2} \bigl ) = 1, \qquad &
   \qquad  \bigl ({k^2\over h_t^2} \bigl ) = 0 ;\cr
 \bigl ({\lambda^2\over h_t^2} \bigl ) = 0, \qquad &
   \qquad  \bigl ({k^2\over h_t^2} \bigl ) = 1 ;\cr
 \bigl ({\lambda^2\over h_t^2} \bigl ) = {3\over 4}, \qquad &
   \qquad  \bigl ({k^2\over h_t^2} \bigl ) = {3\over 8}. \cr
}\eqnlbl{\fixpts} $$
Of these fixed points, only the last is infra-red stable. This is
illustrated in Figure 1a, where the paths traced out by a number of
equally spaced points in the $\lambda/ h_t-k/h_t$ plane with $g_i=0$
and $h_t=10$ at the GUT scale are plotted as the energy scale
runs from the GUT scale of $10^{16}$GeV to $10^3$GeV. A point in this
plane will flow rapidly towards the central valley which passes
through all three fixed points, and then more slowly along it to the
stable fixed point. From this figure, it is clear that, regardless of
the high energy values, the low energy values of the couplings are
likely to be somewhere in this one dimensional region, but they may
well be nearer an unstable than a stable fixed point.

So far, however, we have neglected the effects of the large QCD
coupling in equation \RGratios; this may be important as discussed in
[\ref{\bs}].  When we include the effect of gauge couplings in our
numerical analysis, a different picture emerges. Although at high
energy the non-zero QCD coupling is small enough not to make much
difference, and the flow is much as before, at lower energies this
contribution dominates, and the effect is that points flow towards and
along the valley for a short distance before being forced towards the
origin at lower energies. However, it is noticeable that the region of
parameter space to which the couplings flow in this figure is still
one dimensional. This is illustrated in Figure 1b, which again has
$h_t=10$ at the GUT scale.

Both of these plots have been done in the case where the couplings
$\lambda$ and $k$ are large; if they are small ($<1$ at high energy),
then the flow towards the fixed point in the gaugeless limit is much
slower, and with gauge couplings included most points flow rapidly to
the origin remaining within a region bounded by the one dimensional
valley described above and the x- and y-axes, as shown in Figure 1c,
where $h_t=1$ at the GUT scale. Finally, if $\lambda$, $k$ are greater
than $h_t$ at high energies, then they flow very rapidly towards the
origin and end up on the boundary of this region.

In summary, for fixed $h_t$ the renormalisation group flows give a
curve in parameter space which can be attained consistent with
triviality from the assumption of large couplings at the GUT scale,
but any point between this curve and the origin is also possible given
appropriate initial values of the couplings.

This analysis, with the assumption that the high energy behaviour of
the theory is described by a GUT model, or at least that perturbative
physics continues up to the unification scale of order $10^{16}$GeV,
means that we can find maximum values of $\lambda$ and $k$ consistent
with a given value of $h_t$. If we take $k=0$ (in order to maximise
the low energy value of $\lambda$), it is possible to solve for
$\lambda_{\rm max}$ (the value of $\lambda$ at which it just becomes
non-perturbative at the unification scale) as a function of
$h_t$(1TeV) (or equivalently of $m_t$ and $\tan\beta$). This has been
done [\ref{\eq},\ref{\kkw}], and the results are reproduced in Figure
2.

\num=0
\sec=4
\heading {4 Radiative Corrections }
There have been a number of attempts to calculate radiative corrections
to the tree-level results presented in section 2 [\ref{\eltraub},
\ref{\ellind}, \ref{\el}, \ref{\terV}, \ref{\eqtwo}, \ref{\ekwone},
\ref{\ekwtwo}, \ref{\pandita}]. In the MSSM, the dominant radiative
corrections are those resulting from loops of top quarks and stop
squarks. These corrections have been calculated using a range of
diagrammatic, RG and one-loop effective
potential techniques.
Radiative corrections due to the top quark and
stop squark for the upper $2\times 2$ part of
the mass matrix are identical to those of the MSSM if we replace $\mu$
by $\lambda x$, and these have been studied in a simple approximation
[\ref{\ellind}]. The full analytic corrections to other components of
the scalar and pseudo-scalar Higgs mass matrices have also been
calculated [\ref{\el}, \ref{\ekwtwo}, \ref{\pandita}].

Logarithmic effects from loops of light Higgs bosons have been studied
using a full RG analysis below a hard SUSY breaking scale [\ref{\terV},
\ref{\ekwone}].  However these analyses may be criticised on two counts.
Firstly, the assumed spectrum  of degenerate sparticles at some SUSY
scale and Higgs bosons degenerate with the top quark at some lower scale
is over-simplified. Secondly, the RG analysis fails to pick up
non-logarithmic radiative corrections depending on soft trilinear
couplings. These corrections are known to be significant in the squark
sector, and the corresponding corrections from the Higgs sector may also
be important. It is possible to complicate the RG approach to take
account of some of these effects, by systematically decoupling heavy
particles below their mass thresholds, and introducing finite shifts in
boundary conditions, but the elegance of the RG approach is then lost
[\ref{\haber}]. For the cases where the effects of finite diagrams are
of interest, the simplest approach is to perform a one-loop effective
potential calculation. In reference [\ref{\ekwtwo}] we considered a
hybrid approach in which squark corrections to all Higgs boson masses
were calculated in the framework of the one-loop effective potential,
and these corrections were grafted on to our previous RG analysis
[\ref{\ekwone}] involving light Higgs bosons and a degenerate top quark.

In this section we shall abandon the RG approach completely and perform
a full one-loop effective  potential analysis of radiative corrections
involving the top quark, stop squarks, Higgs bosons, and Higgsinos.
We shall introduce the use of the effective potential for calculating
radiative corrections to mass matrices, present corrections due to
the top quark and stop squarks, and discuss the
calculation of Higgs and Higgsino corrections.

\smallheading{4.1 One-Loop Effective Potential}
The full one-loop corrected scalar potential is given by
$$
V_1=V_0+\Delta V_1,
\eqnlbl{\cwpotl} $$
where $V_0$ is the tree level potential, and $\Delta V_1$ is
$$
\Delta V_1 = {1 \over 64 \pi^2} {\rm Str} {\cal M}^4
                   \Bigl (\log{{\cal M}^2\over \mu^2} - {3\over 2} \Bigr ).
\eqnlbl{\deltaV} $$
$\mu$ is the $\overline{\hbox{MS}}$ renormalisation scale, and the
supertrace is a trace over all fields which couple through the mass
matrix and includes a factor $(-1)^{2J} (2J+1)$ so that a Weyl fermion
acquires a factor -2, a real scalar a factor 1, and we must remember
appropriate colour and flavour factors. ${\cal M}^2$ is the field
dependent mass-squared matrix, in which the fields left after
differentiating $V_0$ to obtain ${\cal M}^2$ are not replaced by their
vevs.

The one-loop minimisation conditions are, of course, ${\partial V_1 \over
\partial \phi_i} = 0$, $i=1,2,3$. After replacing the fields by their
vevs, we have explicitly
$$\eqalign{
2m_1^2\nu_1 + 2 \lambda_1\nu_1^3 + 2 (\lambda_3+\lambda_4)\nu_1\nu_2^2
 + 2\lambda_5x^2\nu_1 + 2\lambda_7x^2\nu_2 - m_4x\nu_2
 + \sqrt{2}\left.{\partial\Delta V_1\over\partial\varphi_1}
         \right\vert_{vevs}  = & 0 \cr
2m_2^2\nu_2 + 2 \lambda_2\nu_2^3 + 2 (\lambda_3+\lambda_4)\nu_1^2\nu_2
 + 2\lambda_6\nu_2x^2 + 2\lambda_7\nu_1x^2 - m_4\nu_1x
 + \sqrt{2}\left .{\partial\Delta V_1\over\partial\varphi_2}
         \right\vert_{vevs}  = & 0 \cr
2m_3^2x + 2 \lambda_5\nu_1^2x + 2\lambda_6\nu_2^2x
 + 4\lambda_7 \nu_1\nu_2x + 4\lambda_8 x^3
 - 2m_4\nu_1\nu_2 - 2 m_5 x^2
 + \sqrt{2}\left.{\partial\Delta V_1\over\partial\varphi_3}
         \right\vert_{vevs}  = & 0 \cr
}\eqnlbl{\mincond}
$$
from which we can see that
$$
m_i^2 = \left. m_i^2 \right|_{TL} - {1\over \nu_i \sqrt{2}}
  \left.{\partial \Delta V_1 \over \partial \phi_i}\right\vert_{vevs}
\eqnlbl{\newms}
$$
where $m_i^2\vert _{TL}$ are the values of $m_i^2$ in terms of the vevs
as evaluated at tree-level. Having thus supplied the vevs, we extremise
the potential by choosing the soft masses to satisfy equation \newms.
This does not guarantee that this extremum is a minimum of the
potential; this can only be done by calculating the physical Higgs
spectrum and ensuring that all the masses squared  are positive.

The minimum so constructed need not be the global minimum of the Higgs
potential. Thus we examine the potential containing these (radiatively
corrected) mass parameters $m_i^2$ to determine whether alternative
minima with zero vevs for one or more of the scalar fields
$\varphi_1$, $\varphi_2$, and $\varphi_3$ exist. In fact the case
where only one of these is zero does not occur for any of our choices
of parameters.  We test such possible minima by comparing the value of
the potential, including top and stop corrections only for simplicity,
to discover whether these alternative minima are preferred; if so we
discard this point in parameter space.

The radiatively corrected matrices for the CP-even, CP-odd and charged
Higgs scalars may be calculated in a straightforward manner. In the
basis defined by \phidefs, the correction to the mass-squared matrix is
$$
{\delta\cal M}_{ij}^2
  = {\partial^2 \Delta V_1 \over \partial \phi_i \partial \phi_j}.
\eqnum
$$
In fact, this is only approximately true due to Higgs self--energy
corrections; these are expected to be small for the lightest Higgs
bosons [\ref{\brignole}].  However we should recall that the formula for
${\cal M}^2$ given in equation \fullMsq\ contains implicit dependence on
the mass parameters $m_i^2$, and so will involve dependence on the first
derivative of  $\Delta V$. Thus for example the radiative corrections to
the scalar mass-squared matrix $M^2$ whose tree-level value is given in
equation \scalmatrix\ are of form
$$
\delta M_{ij}^2 = \left. \left(
{\partial^2 \Delta V_1\over \partial \phi_i \partial \phi_j}
-{1\over \nu_i \sqrt{2}}{\partial\Delta V_1\over \partial \phi_j}
 \delta_{ij} \right) \right|_{vevs}.
\eqnlbl{\shiftmass}
$$
The last term represents the shift in the mass matrix due to the
radiative corrections to the minimisation conditions.

\smallheading{4.2 Squark and Top Corrections}
The corrections to the mass matrices due to top quark and stop squark
loops have been calculated elsewhere [\ref{\ellind}, \ref{\el},
\ref{\ekwtwo}, \ref{\pandita}], so we simply give the results, in the
notation of [\ref{\ekwtwo}], without further explanation. The field
dependent top quark and stop and sbottom squark mass matrices are given
by
$$
M^2_{\rm top} = h_t^2 ( \vert H_2^0 \vert^2 + \vert H_2^+ \vert^2 )
\eqnlbl{\topmass}
$$
$$
M^2_{\rm sq} = \left(  \matrix{
          m_Q^2 + h_t^2 \vert H_2^0 \vert^2             &
          h_t \lambda N H_1^0 + h_tA_t\bar H_2^0        &
          -h_t^2\bar H_2^0 H_2^+                        &
          0                                             \cr
          h_t \lambda\bar N \bar H_1^0 + h_tA_tH_2^0    &
          m_T^2 + h_t^2( \vert H_2^0 \vert^2 + \vert H_2^+ \vert^2 ) &
          h_t \lambda \bar N \bar H_1^- - h_tA_t H_2^+  &
          0                                             \cr
          -h_t^2 H_2^0 \bar H_2^+                       &
          h_t \lambda N H_1^- - h_tA_t \bar H_2^+       &
          m_Q^2 + h_t^2 \vert H_2^+ \vert^2             &
          0                                             \cr
          0 & 0 & 0 & m_B^2 \cr  }
                                 \right ),
\eqnlbl{\stopmass}
$$
in the basis
$\{ \tilde{t_L},\bar{\tilde{t_R^c}},\tilde{b_L},\bar{\tilde{b_R^c}} \}$,
and a bar denotes complex conjugation. The sbottom squarks appear because
they contribute to the charged Higgs mass. One eigenvalue of \stopmass\ is
field independent and may be discarded. Notice that the form of this
equation implies that
$$
A_t+\lambda x \cot\beta\leq
   {\vert m_{\tilde t_2}^2-m_{\tilde t_1}^2 \vert \over 2 h_t\nu_2}
\eqnlbl{\Aplus}
$$
Here $m_{\tilde t_1}$ and $m_{\tilde t_2}$ are the stop squark mass
eigenvalues.

After considerable algebra we obtain the following corrections to the
Higgs mass matrices. The correction to the CP-odd mass-squared matrix
${\tilde{M}}^2$ in equation \pseudmatrix\ is given by
$\delta{\tilde{M}}^2$, where
$$
\delta {\tilde{M}}^2 = \left( \matrix{
   \tan \beta &    1       &   {\sin \beta\over r} \cr
       1      & \cot \beta &   {\cos \beta\over r} \cr
{\sin\beta\over r} & {\cos\beta\over r} &{\sin\beta\cos\beta\over r^2} \cr
} \right) \Delta^2,
\eqnum $$
with
$$
\Delta^2 = {3\over 16 \pi^2} h_t^2 (\lambda x) A_t
                 f(m_{\tilde t_1}^2,m_{\tilde t_2}^2),
\eqnum$$
and the function $f$ is defined by
$$
f(m_{\tilde t_1}^2,m_{\tilde t_2}^2) =
      {1\over m_{\tilde t_2}^2-m_{\tilde t_1}^2 }
     \Bigl (
       m_{\tilde t_1}^2 \log\bigl({m_{\tilde t_1}^2\over \mu^2}\bigr)
     - m_{\tilde t_2}^2 \log\bigl({m_{\tilde t_2}^2\over \mu^2}\bigr)
     - m_{\tilde t_1}^2
     + m_{\tilde t_2}^2
        \Bigr )
\eqnum$$
The correction to the CP-even mass-squared matrix $M^2$ in equation
\scalmatrix\ is given by $\delta M^2$, where
$$
\delta M^2 = \left( \matrix{
  \Delta_{11}^2 & \Delta_{12}^2 & \Delta_{13}^2 \cr
  \Delta_{12}^2 & \Delta_{22}^2 & \Delta_{23}^2 \cr
  \Delta_{13}^2 & \Delta_{23}^2 & \Delta_{33}^2 \cr
} \right) + \left( \matrix{
   \tan \beta &    -1       &   -{\sin \beta\over r} \cr
      -1      & \cot \beta &   -{\cos \beta\over r} \cr
-{\sin\beta\over r}&-{\cos\beta\over r} &{\sin\beta\cos\beta\over r^2}\cr
} \right) \Delta^2,
\eqnlbl{\scalarcor} $$
where the $\Delta_{ij}^2$ are given by
$$\eqalign{
\Delta_{11}^2 =\, & {3\over 8 \pi^2} h_t^4 \nu_2^2 (\lambda x)^2
 \Bigl(
  {A_t+\lambda x\cot\beta \over m_{\tilde t_2}^2 - m_{\tilde t_1}^2}
          \Bigr )^2  g(m_{\tilde t_1}^2,m_{\tilde t_2}^2) \cr
\Delta_{12}^2 =\, & {3\over 8 \pi^2} h_t^4 \nu_2^2 (\lambda x)
 \Bigl(
  {A_t+\lambda x\cot\beta \over m_{\tilde t_2}^2 - m_{\tilde t_1}^2}
          \Bigr )
 \Bigl(
       \log \bigl({ m_{\tilde t_2}^2 \over m_{\tilde t_1}^2 }\bigr ) \cr
 &\qquad \qquad
  +{A_t(A_t+\lambda x\cot\beta)\over m_{\tilde t_2}^2 - m_{\tilde t_1}^2}
        g(m_{\tilde t_1}^2,m_{\tilde t_2}^2) \Bigr ) \cr
\Delta_{13}^2=\, &{3\over 8 \pi^2}h_t^4\nu_2^2 (\lambda x)(\lambda \nu_1)
 \Bigl(
  {A_t+\lambda x\cot\beta \over m_{\tilde t_2}^2 - m_{\tilde t_1}^2}
          \Bigr )^2   g(m_{\tilde t_1}^2,m_{\tilde t_2}^2) \cr
& \qquad
   - {3\over 8 \pi^2}h_t^2\nu_2^2 (\lambda x)(\lambda \nu_1)
        f(m_{\tilde t_1}^2,m_{\tilde t_2}^2) \cr
\Delta_{22}^2 =\, & {3\over 8 \pi^2} h_t^4 \nu_2^2
 \Bigl(
       \log \bigl({ m_{\tilde t_1}^2 m_{\tilde t_2}^2 \over m_t^4 }\bigr)
 + {2A_t(A_t+\lambda x\cot\beta)\over m_{\tilde t_2}^2 - m_{\tilde t_1}^2}
       \log\bigl ( {m_{\tilde t_2}^2 \over m_{\tilde t_1}^2 }\bigr )
          \Bigr ) \cr
 & \qquad +{3\over 8 \pi^2} h_t^4 \nu_2^2
 \Bigl(
  {A_t(A_t+\lambda x\cot\beta) \over m_{\tilde t_2}^2 - m_{\tilde t_1}^2}
          \Bigr )^2    g(m_{\tilde t_1}^2,m_{\tilde t_2}^2) \cr
\Delta_{23}^2 = \, & {\cos\beta\over r} \Delta_{12}^2 \cr
\Delta_{33}^2 = \, & {3\over 8 \pi^2} h_t^4 \nu_2^2 (\lambda\nu_1)^2
 \Bigl(
  {A_t+\lambda x\cot\beta \over m_{\tilde t_2}^2 - m_{\tilde t_1}^2}
          \Bigr )^2  g(m_{\tilde t_1}^2,m_{\tilde t_2}^2) \cr
}\eqnum$$
and the function $g$ is defined by
$$
g(m_{\tilde t_1}^2,m_{\tilde t_2}^2) =
      {- 1\over m_{\tilde t_2}^2-m_{\tilde t_1}^2 }
     \Bigl (
       (m_{\tilde t_1}^2+m_{\tilde t_1}^2)
            \log\bigl({m_{\tilde t_2}^2\over m_{\tilde t_1}^2}\bigr)
     + 2m_{\tilde t_1}^2
     - 2m_{\tilde t_2}^2
        \Bigr )
\eqnlbl{\gdefine}$$
Finally, the correction to the charged mass-squared matrix $M_c^2$ in
equation \chargedmatrix\ is $\delta M_c^2$, where
$$
\delta M_c^2=
    \left(\matrix{ \tan\beta &     1     \cr
                       1     & \cot\beta \cr} \right )
                      \Delta_c^2
\eqnum$$
and
$$\Delta_c^2 = {3\over 16 \pi^2}
   \Sigma_{i\in \{ \tilde t_1, \tilde t_2, \tilde b_1 \} }
       m_i^2 \Bigl ( \log\bigl( {m_i^2\over m_t^2} \bigr ) -1 \Bigr )
       {\partial^2 m_i^2\over\partial H_1^- \partial H_2^+ }
        \vert_{vevs}
\eqnum $$
$$\eqalign{
{\partial^2 m_{\tilde t_1}^2\over\partial H_1^- \partial H_2^+ }
  \vert_{vevs}  = \, &
   - {h_t^4\nu_2^2(\lambda x)^2\cot\beta\over
         (m_{\tilde t_1}^2-m_{\tilde t_2}^2)
         (m_{\tilde t_1}^2-m_{\tilde b_1}^2)}
   - {h_t^2(\lambda x)A_t\over
         (m_{\tilde t_1}^2-m_{\tilde t_2}^2)} \cr
{\partial^2 m_{\tilde t_2}^2\over\partial H_1^- \partial H_2^+ }
  \vert_{vevs}  = \, &
   - {h_t^4\nu_2^2(\lambda x)^2\cot\beta\over
         (m_{\tilde t_2}^2-m_{\tilde b_1}^2)
         (m_{\tilde t_2}^2-m_{\tilde t_1}^2)}
   + {h_t^2(\lambda x)A_t\over
         (m_{\tilde t_1}^2-m_{\tilde t_2}^2)} \cr
{\partial^2 m_{\tilde b_1}^2\over\partial H_1^- \partial H_2^+ }
  \vert_{vevs}  = \, &
   - {h_t^4\nu_2^2(\lambda x)^2\cot\beta\over
         (m_{\tilde b_1}^2-m_{\tilde t_1}^2)
         (m_{\tilde b_1}^2-m_{\tilde t_2}^2)} \cr
}\eqnum$$

\smallheading{4.3 Higgs and Higgsino Corrections}
Calculation of the corrections due to Higgs loops is much more
cumbersome because the field dependent Higgs mass-squared matrix in
equation \fullMsq, used to calculate $\Delta V_1$ in equation \deltaV,
is $10\times 10$.  Since it is impossible to determine the eigenvalues
of general matrices larger than $4\times 4$ analytically, resort to
numerical techniques is essential. As is clear from \shiftmass, we are
interested in the first and second derivatives of $\Delta V_1$ evaluated
at the vevs, and these are given by
$$ \left. {\partial \Delta V_1 \over \partial \phi_i} \right\vert_{vevs}=
\left. {1\over 32\pi^2} m^2_\alpha {\partial m^2_\alpha \over \partial
\phi_i}
\left( \log {m^2_\alpha \over \mu^2} - 1 \right) \right\vert_{vevs},
\eqnum$$
$$
\left. {\partial^2 \Delta V_1 \over
\partial \phi_i \partial \phi_j} \right\vert_{vevs} =
\left. {1\over 32\pi^2}{\partial m^2_\alpha \over \partial \phi_i}
{\partial m^2_\alpha \over \partial \phi_j}
\log {m^2_\alpha \over \mu^2} \right\vert_{vevs}
+
\left. {1\over 32\pi^2} m^2_\alpha
{\partial^2 m^2_\alpha \over \partial \phi_i \partial \phi_j}
\left( \log {m^2_\alpha \over \mu^2} - 1 \right) \right\vert_{vevs},
\eqnum$$
where $\{m^2_\alpha \}$ is the set of the 10 eigenvalues of ${\cal
M}_{ij}^2$, and we implicitly sum over $\alpha$. It is straightforward
numerically to obtain $m^2_\alpha \vert_{vevs}$ from ${\cal M}_{ij}^2
\vert_{vevs}$, and since the first and second derivatives of
$m^2_\alpha$ evaluated at the vevs may also be obtained numerically,
the problem reduces to a routine numerical task.

Similarly, since the Higgsino mass matrix is $5 \times 5$, Higgsino
loop contributions must be calculated numerically as above. We have
neglected the mixing with gauginos for simplicity, as in this way we
can eliminate two soft mass parameters. Moreover, if we include
gaugino corrections, then we should include all gauge boson
corrections for consistency, particularly since there are typically
cancellations between particles and their superpartners, as we shall
see later. The field dependent Higgsino mass term, with Higginos
written as 4-component Majorana spinors, in the basis
$\psi=(\tilde{H_1^0},\tilde{H_2^0},\tilde{N},
\tilde{H_1^-},\tilde{H_2^+})^T$,
is given by $\half\overline{\psi} M_{\tilde H}\psi$, with $M_{\tilde
H}$ defined by
$$
M_{\tilde H} = {\rm Re} M' - i \gamma_5 {\rm Im} M' ,
\eqnum$$
where
$$
M'= \left( \matrix{
     0        &  \lambda N    &  \lambda H_2^0  &    0     &     0 \cr
 \lambda N    &     0         &  \lambda H_1^0  &    0     &     0  \cr
\lambda H_2^0 & \lambda H_1^0 & -2k N & -\lambda H_2^+ & -\lambda H_1^- \cr
     0        &     0         & -\lambda H_2^+   &    0     & -\lambda N \cr
     0        &     0         & -\lambda H_1^-   & -\lambda N &    0  \cr }
     \right).
\eqnum$$
The contribution to $\Delta V_1$ from Higgsinos is then given by
$$
-{1\over 128\pi^2} {\rm Tr} \left\{ (M_{\tilde H}M_{\tilde H}^\dagger)^2
  \left[\log\left( {M_{\tilde H}M_{\tilde H}^\dagger\over \mu^2} \right)
- {3\over 2} \right]
\right\},
\eqnum$$
where the trace is over Dirac as well as internal indices. Note that
there is a relative factor of 2 between this result and that of
[\ref{\sher}] because we are working with Majorana spinors.

The numerical calculations are of course very computationally
intensive; however, numerical inaccuracies should generally be less
than about 1GeV, particularly in regions of parameter space where all
of the particles are light. It appears unlikely that the inclusion of
two-loop effects or removing any of the approximations in the
calculations would give corrections significant relative to our
uncertainty in the parameters.

\num=0
\sec=5

\heading{5 The Bound on the Lightest CP-even Higgs}
An upper bound on the lightest neutral CP-even scalar in the NMSSM
may be obtained from the real symmetric $3\times 3$ neutral scalar mass
squared matrix by using the fact that its minimum eigenvalue
is less than or equal to the minimum eigenvalue of its
upper $2 \times 2$ block with the result [\ref{\dl}]
$$
{m_{h}}^2\leq {M_Z}^2 +
(\lambda ^2v^2-{M_Z}^2)\sin^22\beta.
\eqnum
$$
The upper bound on $m_h$ is determined by  the maximum value of $\lambda
(M_{SUSY})$, henceforth  denoted $\lambda _{max}$.
As discussed in section 3, the value of $\lambda _{max}$ is obtained by
solving the SUSY RG equations for the Yukawa couplings $h_t$, $\lambda$
and $k$ in the region  $M_{SUSY}=1$ TeV to $M_{GUT}=10^{16}$ GeV
[\ref{\eq},\ref{\bs}]. There we found that for
$h_t(M_{SUSY})=0.5-1.0$,   ${\lambda}_{max}=0.87-0.70$ and for
$h_t(M_{SUSY}) \rightarrow 1.06$,\footnote*{ $h_t(M_{SUSY})\leq 1.06$ is
the triviality bound, which, together with $m_t=h_t(m_t)v\sin \beta$,
where $h_t(m_t)\leq 1.12$, implies the bound $m_t \leq 195$ GeV.}
${\lambda}_{max}\rightarrow 0$ (with $k=0$ always).

We now wish to evaluate the bound including all the radiative
corrections.  The parameters on which the radiatively corrected bound
depends are $m_t$, $\sin\beta$, $r$, $\lambda$, $k$, $m_4$, $m_5$,
$A_t$, $m_{\tilde t_1}$, and $m_{\tilde t_2}$ (sbottom masses do not
affect the bound, because they only occur in the radiative corrections
to the charged Higgs mass as discussed in reference [\ref{\ekwtwo}]).
We adopt the following approach to these variables. $\lambda$ will,
for given $m_t$ and $\sin\beta$, be given by its maximum value
consistent with remaining perturbative, $\lambda_{\rm max}$. Strictly
speaking $k$ and $m_5$ should then be zero, but since this results in
an axion and thus an unacceptable spectrum, we shall instead use the
value 0.1 for $k$. In fact sensitivity of the bound (and of
$\lambda_{\rm max}$) to $k$ and $m_5$ is negligible when they are both
small (0.1 is sufficiently small for $k$, while small for $m_5$ means
of order tens of GeV). This should however be borne in mind when we
come to discuss the spectrum, where the lightest scalar is quite
sensitive to these variables.  Of the remaining parameters, $m_t$ is
taken as an input parameter, as is the heavier stop mass $m_{\tilde
t_2}$.

The bound also depends on $r$ which may be saturated by requiring that
the larger eigenvalue of the upper $2\times 2$ block is also an
eigenvalue of the whole $3\times 3$ matrix. This can be done by
performing a unitary transformation consisting of an upper $2 \times
2$ unitary matrix, parametrised by an angle $\theta$, together with 33
element equal to unity and all other elements equal to zero.  The
condition that the 13 element of the transformed matrix be zero leads
to the constraint
$$
\cos\theta M^2_{13} + \sin\theta M^2_{23} = 0
\eqnum
$$
where $M^2_{ij}$ are components of the mass matrix in \scalmatrix.
While this guarantees that the bound must be saturated, it is possible
that there is another lighter mass eigenstate consisting of a mixture
of singlet and doublet states. This state will certainly depend
strongly on the values of $k$ and $m_5$.  Examples of this will be
discussed in section 6.

We have thus succeeded in reducing our parameter space to four
variables, $\sin\beta$, $m_{\tilde t_1}$, $A_t$, and $m_4$.  These
variables will all be determined numerically using the Nelder-Mead
simplex method [\ref{\numrec}] to maximise the bound as a function of
the parameters, imposing the constraint \Aplus.

At this point it is worth recalling that in the MSSM the relevant
parameters may be taken to be $\tan\beta$ and $m_{\rm c}$.  Using the
fact that in the minimal model the bound is always maximised for
$\sin\beta=1$, the bound is then equal to the 22 element of
the mass matrix $M^2+\delta M^2$ in \scalmatrix\ and \scalarcor, and
is given by
$$\eqalign{
  m_h^2\le M_Z^2 &
+ {3\over 4\pi^2}h_t^4\nu^2
          \log\left({m_{\tilde t_2}^2\over m_t^2 }\right )\cr
&+ {3\over 4\pi^2}h_t^4\nu^2
   \bigl ( {A_t^2\over m_{\tilde t_2}^2-m_{\tilde t_1}^2 } - {1\over 2}
   \bigr)
   \log\left( {m_{\tilde t_2}^2\over m_{\tilde t_1}^2 }\right )\cr
&+ {3\over 8\pi^2}h_t^4\nu^2
   \bigl ( {A_t^4\over (m_{\tilde t_2}^2-m_{\tilde t_1}^2)^2 } \bigr)
    g(m_{\tilde t_1}, m_{\tilde t_2})\cr
}\eqnlbl{\MSSMRC} $$
where the function $g$ is defined in \gdefine. Since the coefficient
for the $A_t^4$ term is always negative \footnote*{ Note that $g\le 0$
and $g/(m_{\tilde t_2}^2-m_{\tilde t_1}^2)^2$ remains finite as
$m_{\tilde t_1}$ tends to $m_{\tilde t_2}$.} the bound will be
maximised for some non-zero value of $A_t^2$.  The constraint \Aplus\
must also be implemented so that value of $A_t$ will not always reach
the value which maximises the polynomial.  Finally, we note that if we
allow $m_{\tilde t_2}$ to become comparable in size to the top mass,
then the coefficient of the second logarithm in equation \MSSMRC\ will
become negative, and so the bound will be maximised when the two
squarks are degenerate; this typically occurs only for very large top
mass, even if we allow $m_{\tilde t_2}$ to be as low as 250GeV.

Let us now proceed to the discussion of our results for the bound in
the NMSSM (and MSSM for comparison) with all radiative corrections
included.  Figure 3 shows the values of $\sin\beta$, $h_t$, and
$\lambda$ for which the bound is saturated as a function of the top
mass, derived as described above. Although this figure is for
$m_{\tilde t_2}=500$GeV, the results which can be derived for other
values of $m_{\tilde t_2}$ are virtually identical. This figure is
very similar to Figure 1 of reference [\ref{\ekwone}], which was,
however, derived without the inclusion of squark effects, as discussed
there.

Values of the bound as a function of the top mass with the heavier
stop mass $m_{\tilde t_2}$ taken to be 500 and 1000 GeV are shown in
Figure 4. It is clear from this graph that for low top mass the
dependence on the stop mass is very slight (as should be obvious, as
here $h_t$ is small), while for larger top masses this dependence is
far greater, as the radiative corrections are very large. At tree
level the decreasing value of $\lambda$ means that the bound is a
monotonically decreasing function of $m_t$, but for large enough
$m_{\tilde t_2}$ the bound actually starts increasing again with
increasing top mass (both the top corrections which have a
$\log({m_{\tilde t}\over m_t})$ dependence and the stop corrections
can become large only if $m_{\tilde t_2}$ is large). Numerically the
bound is around 150 to 155GeV, and is maximised for small top mass;
but for top masses near the triviality bound and stop masses up to
1TeV this same value is also approached. The radiative corrections to
the bound from Higgs and Higgsino loops are relatively small, of order
a few GeV, and since they are of opposite sign (including Higgs loops
decreases the bound, while including Higgsinos increases it) the total
effect is typically only one or two GeV. In general the optimum value
of $A_t$ is of order 1.6TeV to 2TeV for $m_{\tilde t_2}=1$TeV and of
order 600GeV to 800GeV for $m_{\tilde t_2}=500$GeV. The dependence of
this optimum value on the top mass is very small, but generally it
decreases as $m_t$ increases. Typically the value of $m_{\tilde t_1}$
which maximises the bound is around a third to three quarters the
value of $m_{\tilde t_2}$.

The MSSM bound for the same values of the stop masses is also shown in
Figure 4 for comparison. In this figure Higgs and Higgsino radiative
corrections are included in addition to the top quark and stop squark
corrections in \MSSMRC.  Note that, unlike the case in the NMSSM, the
MSSM bound is a monotonically increasing function of top mass, and
that it approaches the NMSSM bound for sufficiently large $m_t$.

The bound is also presented in Figures 5 but here as a function of
$m_{\tilde t_1}$, the lighter stop mass, for $m_t=150$GeV and
$m_{\tilde t_2}=$500GeV and 1000GeV respectively. As can be seen from
these graphs, there is a gradual rise in the bound until it reaches
the maximum before falling off more rapidly. The very rapid drop in
the bound as $m_{\tilde t_1}$ approaches $m_{\tilde t_2}$ is caused by
the constraint \Aplus, which forces $A_t+\lambda x\cot\beta$ to zero
in this region. The most interesting feature of these figures is that
they suggest that an arbitrarily chosen value of the lighter stop mass
is not unlikely to give a value of the bound close to the maximum
possible, since the variation of the bound with $m_{\tilde t_1}$ is
relatively small over much of its range.

\num=0
\sec=6

\heading{ 6 Higgs Boson Phenomenology}

\noskipsmallheading{6.1 Preliminary Remarks }
In this section we shall explore some of the phenomenological
implications of our results. The qualitative features of certain aspects
of our discussion are well known at the tree-level [\ref{\NMSSM}];
however we are now in a position to re-examine some of these issues in
the light of our treatment of radiative corrections.

We have already emphasised that the lightest CP-even neutral scalar $h$
(the analogue of the Higgs boson of the standard model) may be heavier
in the NMSSM than in the MSSM, although as $m_t\rightarrow 190GeV$ the
two bounds coalesce as shown in Figure~4. However from a pragmatic
standpoint the key question is whether $m_h$ can exceed $M_Z$ and so
evade discovery at LEPII. This can happen in both the MSSM and the
NMSSM due to radiative corrections and thus failure to observe $h$ at LEPII
cannot be interpreted as evidence for favouring the NMSSM over the MSSM.
Also we observe that the NMSSM bound $m_h < 150GeV$ means that if $h$ is not
found at LEPII it must necessarily be in the  intermediate mass region
$M_Z-2M_W$ ({\it i.e.,} too heavy to be produced at LEPII but too light
to decay into pairs of Ws or Zs) in both models. Thus the  upper bound
on $m_h$ may not be particularly helpful in enabling us to distinguish
between the MSSM and the NMSSM.

We can turn the argument of the preceding paragraph around and ask
whether the observation  of $h$ would enable the MSSM and NMSSM to be
distinguished. This question is analogous to that of distinguishing
between the minimal standard model (MSM) and the MSSM from an
observation of the Higgs boson, and has been widely considered
[\ref{\hhguide},\ref{\roger}]. The answer depends in part on measuring
the $ZZh$ coupling which contains a factor of
$R_{ZZh}=sin(\beta-\alpha)$ relative to the usual standard model
coupling, where $\alpha$ is a mixing angle which results from
diagonalising the scalar $2\times 2$ matrix. Although $R_{ZZh}$ may be
small, the $ZAh$ coupling contains a factor of
$R_{ZAh}=cos(\beta-\alpha)$ so that both couplings cannot simultaneously
be small. In the NMSSM the situation is not so simple since the $ZZh$
coupling is derived from diagonalising the scalar  $3\times 3$ matrix
and so $R_{ZZh}$ is more complicated [\ref{\NMSSM}]. Similarly the $ZAh$
coupling, where $A$ is the lightest pseudoscalar, in the NMSSM will
contain a more complicated factor $R_{ZAh}$ [\ref{\NMSSM}].

Given that in the NMSSM physical Higgs boson spectrum  there are three
CP-even scalars and two CP-odd pseudoscalars, one more in each case than
in the MSSM, it may seem at first sight that the Higgses of the NMSSM,
being more numerous, are therefore easier to discover. Unfortunately
this is not so. The only reason why there are more neutral states is due
to the complex singlet $N$, whose scalar and pseudoscalar components mix
with the neutral components of $H_1$ and $H_2$. Since $N$ has no gauge
couplings, this has the effect of diluting the couplings of the neutral
Higgs particles: there are more of them but they all couple more weakly,
making them harder to produce. However the dilution of the neutral Higgs
couplings may be the key to distinguishing between the NMSSM and the
MSSM, since in the NMSSM there is the possibility of a very light Higgs
boson waiting to be discovered with higher statistics data from LEPI.
This is an intriguing possibility, unique to the NMSSM, because although
it is possible to have  $R_{ZZh}=sin(\beta-\alpha)\approx 0$ in the
MSSM, this is always accompanied by $R_{ZAh}=cos(\beta-\alpha)\approx 1$
and since small $m_h$ is associated with small $m_A$ the light decoupled
Higgs scenario in the MSSM is ruled out. Thus the discovery of a light
weakly coupled Higgs boson would be exactly the sort of signature which
would suggest the NMSSM.

The charged scalars remain unaffected by the presence of the additional
singlet, and their gauge and matter couplings in the NMSSM are identical
to those of the MSSM. However in contrast to the MSSM bound $m_c>M_W$
(in the absence of radiative corrections), in the NMSSM we may have
$m_c<M_W$. There will however be a constraint on how light the charged
scalars can be since they will give a contribution to $b\rightarrow
s\gamma$ via a penguin diagram involving an intermediate top quark line
[\ref{\bsg}]. Such contributions may be partially or completely
cancelled by other SUSY diagrams [\ref{\barb}].  It would therefore be
prudent to keep an open mind on the existence of light charged scalars
in the range $m_c=45GeV-M_W$ which would be accessible to LEPII. We
emphasise that this range is not consistent with the MSSM and so
the discovery of charged Higgs bosons at LEPII would be,
if not a smoking gun of the NMSSM,
then at least one which is loaded and cocked.

The Higgs boson mass spectrum of the MSSM at tree level may be
expressed in terms of the parameter set $\{ m_A,\tan\beta\}$, so that
at tree-level its properties may be expressed as contour plots in the
$m_A,\tan\beta$ plane [\ref{\roger}].  In the NMSSM, the Higgs boson
mass spectrum may be expressed in terms of the parameter set
$\{\lambda,k,m_c,A_k,\tan\beta,r\}$. These masses are conventionally
plotted as a function of $m_c$ for a particular choice of
$\lambda,k,A_k,\tan\beta,r$. There are thus four additional parameters
in the NMSSM which may be taken to be $\lambda,k,A_k$ and $r$ which
are simply not present in the MSSM. For these four additional
parameters, reference [\ref{\NMSSM}] uses $\lambda=0.87$, $k=0.63$
corresponding to an approximate fixed point, and varies $A_k$ over
allowed ranges to produce bands of Higgs boson masses, for three
different $r$ values of 0.1,1,10. Although it is clear from Figure~1a
that there is indeed a fixed point in the gaugeless limit, it is
equally clear from Figures 1b and 1c that with the QCD and other gauge
couplings switched on this fixed point is washed away. Instead there
seems to be a fringe of preferred values of $\lambda$, $k$ roughly
corresponding to a quarter-circle in the $\lambda$-$k$ plane.

If one is tempted to despair at the six dimensional parameter space of
the NMSSM compared to the two dimensional parameter space of the MSSM,
it is worth remembering that when radiative corrections are taken into
consideration things get much worse. In addition to $m_A$ and
$\tan\beta$, the MSSM Higgs sector depends on five additional
parameters $m_t$, $\mu$, $m_{\tilde t_1}$, $m_{\tilde t_2}$ and $A_t$,
even neglecting the bottom quark Yukawa coupling. Actually the NMSSM
fares slightly better since with radiative corrections due to top
quarks and squarks included there are only four additional parameters
$m_t$, $m_{\tilde t_1}$, $m_{\tilde t_2}$ and $A_t$ since there is no
$\mu$ parameter. The Higgs and Higgsino radiative corrections do not
introduce any further parameters since their tree-level masses, which
are used to generate the one-loop effective potential, are determined
in terms of the tree-level parameters. This would of course not be the
case if we had included the full effects of Higgsino-gaugino mixing,
introducing extra gaugino mass paramters. One way to reduce the number
of parameters is to assume universal soft parameters at the GUT scale
and this has been done in the NMSSM [\ref{\NMSSM}, \ref{\eltwo}]. Here
we shall not restrict the soft parameters in this way, and instead
attempt a crude exploration of the full parameter space without any
constraints from the GUT scale apart from the usual SUSY desert
assumption that no couplings blow up below $M_{GUT}$.

We shall  plot Higgs boson masses as a function of $m_c$ over the
regions of $m_c$ which correspond to all Higgs boson masses being
greater than zero at tree-level. This turns out to be a non-trivial
restriction, usually (but not always) due to the lightest scalar mass
$m_h$ diving to zero at the edges of the region. Although radiative
corrections tend to expand this region, since typically they serve to
increase Higgs boson masses, we use the tree-level restriction since it
is the tree-level mass matrices which are used in the calculation of the
one-loop effective potential. Charge breaking in the Higgs  sector is
trivial to check for since it just corresponds to $m_c^2$ becoming
negative. The possibility of squark and slepton vevs as discussed
earlier will not concern us here. However we search for
alternative vacua in which  one or more of $v_1, v_2, x$ take zero
values corresponding to a deeper minimum of the (radiatively corrected)
potential, as discussed in section 4. This can be an important
constraint on the allowed range of $m_c$.

\smallheading{6.2 Higgs Boson Masses and Couplings}
With the preliminaries over we now embark on our exploration of the ten
dimensional parameter space which characterises the Higgs boson spectrum
of the NMSSM.  Clearly a full survey of Higgs boson masses over all of
parameter space is not feasible. The best one can do is to define a set
of baseline parameters and explore the effect of varying each of the
parameters in turn. Our baseline parameters are
$$\eqalign{
\lambda=0.65 \qquad k=0.1 \qquad A_k=&0
    \qquad \tan\beta=1.7 \qquad r=1 \cr m_t=150\GeV \qquad m_{\tilde
t_1}=150\GeV \quad &\quad m_{\tilde t_2}=500\GeV \qquad A_t=0 \cr
}\eqnlbl{\baseline} $$ We choose these particular values of $\lambda$,
$k$, $\tan\beta$ because, for a top mass of 150GeV, they are
approximately those which saturate the bound on $m_h$ as discussed in
section 5, and are therefore associated with heavier values of the
lightest CP-even scalar.  Of the remaining parameters, $A_k=A_t=0$,
$r=1$, are chosen for simplicity, and $m_t=150GeV$ is selected as a
typical value. The stop masses (with $m_{\tilde t_2}$ chosen by
convention to be the heavier) are again familiar from our discussion
of the bound, and ensure that the condition on the stop eigenvalues in
equation \Aplus\ is always satisfied over our ranges of parameters,
with the top quark mass equal in mass to the lightest stop. In all our
plots we shall take $A_k=0$, in accordance with the baseline parameter
set defined above.  In fact, there is often very little freedom of
choice of $A_k$ since it is clear from the tree-level mass matrices
\scalmatrix\ and \pseudmatrix, that allowing a large positive value of
$m_5$ (recall that $m_5=kA_k$) forces the scalar mass matrix to have a
negative eigenvalue, while allowing a large negative eigenvalue forces
the pseudo-scalar mass matrix to have a negative value.  This is further
compounded by the structure of the potential, which means that for
large positive values of $m_5$ the coefficient of the $x^3$ term is
large and negative, and so the alternate minimum with $x\ne 0$ and
$\nu_1=\nu_2=0$ will be preferred.

Figure~6a shows the Higgs boson masses for the set of parameters
defined in equation \baseline, including radiative corrections due to
loops of top quarks, squarks, Higgs bosons and Higgsinos. Over the
charged Higgs mass range $m_c=200-250GeV$ the lightest scalar mass
(lowest solid) varies between $m_h\approx 50-90GeV$ while the lightest
pseudoscalar mass (lower dashed) is approximately $80$ GeV. For
smaller values of $m_c$ than plotted the lightest scalar mass dives
down to zero, while for larger $m_c$ than plotted vacuum with
$v_i,x=0$ is preferred, and these two constraints thus define the
allowed region. In Figure~6b we display the amplitude of $N$ contained
in the lightest scalar (dots), the next-to-lightest scalar (dot-dash)
and the lightest pseudoscalar (double-dot-dash). It is clear that the
lightest scalar is predominantly $N$, especially near $m_c=235GeV$,
while the next-to-lightest scalar shows the opposite behaviour, with a
very small $N$ component, especially where the lightest scalar is
predominantly $N$. The lightest pseudoscalar is nearly entirely
singlet. Also shown in Figure~6b are the couplings $R_{ZZh}$ (solid)
and $R_{ZhA}$ (dashed) which confirm that for $m_c\approx 235GeV$ the
lightest scalar $h$ and pseudoscalar $A$ effectively decouple. This
behaviour was also noted in reference [\ref{\ekwone}] where it was
pointed out that the next-to-lightest scalar must and does respect the
bound at the point where the lightest scalar decouples. However since
its mass exceeds 100 GeV over the whole region, the next-to-lightest
scalar will not be visible at LEPII. The lightest scalar may be
accessible to LEPI near the left-hand end of the region, however.

As an example of a region of parameter space which is excluded by LEPI,
in Figure~7a we show the Higgs boson mass spectrum for a value of
$r=0.1$, with all the other parameters set equal to their baseline
values as in Figure~6. Smaller $r$ is  associated with smaller charged
Higgs masses, so that the LEPI bound $m_c>45 GeV$ is sufficient to
exclude the left-hand half of the region immediately. The constraint
from $b\rightarrow s\gamma$ may exclude the whole of the region as
discussed above, but  even without this constraint the right hand half
of the region is probably excluded from LEPI Higgs searches, as we now
discuss. The lightest scalar has a mass of around 50 GeV or less, and
the lightest pseudoscalar has a mass of about 20 GeV, and both these
particles have very little singlet component, as shown in Figure~7b. The
heaviest scalar and the other pseudo-scalar primarily consist of
singlet, and are approximately degenerate at around half a TeV, playing
no important phenomenological role at LEP.  However the LEP
complementary searches for $Z\rightarrow Z^{\ast}h$ and $Z\rightarrow
hA$  probably eliminate the entire allowed region of masses, since both
$R_{ZZh}$ (solid) and $R_{ZhA}$ (dash) in Figure~7b are quite sizeable
on the right-hand part of the region. It therefore seems likely that the
whole of the region is either excluded or on the point of being excluded
by LEPI.

In Figure~8a we show the spectrum of Higgs bosons for $r=5$  and the
other parameters set equal to their baseline values. In this example the
lightest scalar $h$ contains very little $N$ component particularly in
the middle part of the range, according to Figure~8b, and couples to the
Z  almost identically to the Higgs boson of the MSM, according to  the
solid line in the figure.  The lightest scalar mass exceeds 100 GeV over
much of the region, and so is inaccessible to LEPII except at the
extreme ends of the range. The lightest pseudoscalar and second lightest
scalar are  both predominantly $N$. The charged scalar mass, and the
heaviest scalar and pseudoscalar masses exceed 1 TeV over the entire
range for this value of $r$. Thus this large $r$ parameter set produces
a Higgs spectrum which will not be easy to study even at LHC/SSC.

All the above plots are for $\tan \beta =1.7$. In Figure~9a we  show the
Higgs masses for $\tan \beta =10$ with all the other parameters set
equal to their baseline values. Comparing Figure~9a to Figure~6a we see
that one of the main effects of choosing a larger value of $\tan \beta$
is substantially to increase the values of $m_c$ over which an allowed
solution exits, as we would expect from the tree-level mass matrix
\chargedmatrix. Thus we see that larger values of $m_c$ are associated
with larger values of $\tan\beta$, as well as with larger values of $r$
as previously noted.  With $m_c$ now in the TeV range
the lightest scalar mass is pulled down to
30-50 GeV although since it consists almost completely of $N$ and
couples weakly to the Z (see Figure~9b), it is not ruled out by current
Higgs searches at LEPI. This is a good example of a light weakly coupled
Higgs boson, characteristic of the NMSSM. High statistics LEP data would
be  required to discover this weakly coupled Higgs boson. Similarly the
lightest pseudoscalar in Figure~9a, although lighter than its
counterpart in Figure~6a, remains decoupled and undetectable according
to Figure~9b. The physically relevant next-to-lightest scalar in
Figure~9a remains above 100 GeV, out of range of LEPII.

Next we turn to the question of the effect of varying $k$ and $\lambda$.
For the above plots we selected $k=0.1$ and $\lambda =0.65$  because
these values served to maximise the lightest scalar mass in our analysis
of the bound in section 5. Now we shall explore the effect of choosing
different values of $k$ and $\lambda$, and in this discussion we shall
be guided by our study of RG flows in section 3. By comparing Figures~1a
and 1b it is clear that the fixed point $\lambda=0.87h_t$ and
$k=0.63h_t$ is blown away by gauge (in particular QCD) effects. Instead
we are left with a preferred fringe in the $(\lambda /h_t)$ and
$(k/h_t)$ plane of Figure~1b roughly corresponding to a circle of radius
$\approx 0.6$. This plot is for large $h_t$ at the GUT scale,
corresponding to $h_t\approx1.1$ at low energy; however Figure~1c
shows  similar behaviour for $h_t=1$ at the GUT scale. Our baseline
parameters have $m_t=150 GeV$ and $\tan\beta=1.7$ corresponding to
$h_t=0.92$ at a scale of 1TeV; this is the value by which the points on
the fringe of Figure~1b must be multiplied to obtain the values of
$\lambda$ and $k$.

Following the discussion of the preceeding paragraph, a central point on
the fringe of  $(\lambda/h_t)=(k/h_t)=0.44$  corresponds to
$\lambda=k=0.4$ and the spectrum with these values for the couplings and
all other parameters set equal to their baseline values is plotted in
Figure~10a. Compared to the baseline values of $\lambda=0.65$, $k=0.1$
in Figure~6a, Figure~10a shows $m_c$ values of 120-200 GeV, smaller by
about 100GeV, with the left hand end of the range determined by the
lightest pseudoscalar mass dropping to zero, and the right hand end by
the lightest scalar mass dropping to zero. The latter has a mass of up
to 80 GeV, and Figure~10b shows that the lightest scalar (dots) contains
little singlet resulting in a SM-type ZZh coupling (solid) in contrast
to Figure~6b.  This puts the lightest scalar within the LEP range, with
an excluded region corresponding to the LEP limit
$m_h< 60\GeV$ advancing to the left. As before, the
lightest pseudoscalar is mainly singlet and so has small physical
couplings (dashes in Figure~10b).  The heavier Higgses are somewhat
lighter than before but out of range of LEP.

Figure~11a shows the Higgs mass spectrum for $\lambda=0.1$ and $k=0.6$,
at the other end of the fringe to the baseline values of Figure~6a.
Note that Figure~11a is  similar to Figure~10a,
but the lightest scalar is lighter at around 30 GeV, and since
it contains very little $N$ and couples in a standard Higgs
like way to the Z (Figure~11b) it is excluded by present Higgs
searches. We conclude that the small $\lambda$, large $k$ region
is not preferred phenomenologically.

Having explored all the regions of the fringe of preferred values of
$\lambda$ and $k$ in Figure~1b, we shall now consider taking these
parameters both to be very small. The RG equations in section 3 show
that $\lambda=k=0$ is an acceptable choice, since if $\lambda$ and $k$
are small at the GUT scale, they may still be small at low energies.
Furthermore, it is known that in the limit $\lambda\rightarrow 0$,
$k\rightarrow 0$,  $r\rightarrow\infty$, with $\lambda r$ and $kr$ held
fixed, the NMSSM Higgs sector reduces to that of the MSSM, so this
region is of intrinsic interest for comparisons to the minimal model. In
Figure~12a we show the Higgs spectrum of the NMSSM for $\lambda=0.1$,
$k=0.1$, $r=5$ with all other parameters set equal to their baseline
values. This spectrum should and does resemble that of the MSSM. For one
thing the charged mass exceeds $M_W$, as it should. Also it is clear
from Figure~12b that the lightest pseudoscalar (double-dot-dash) and the
second lightest scalar (single-dot-dash) can be identified with $N$ to a
good approximation over most of the region. The remaining states of
Figure~12a then correspond to the MSSM states. The lightest scalar of
mass 50 GeV couples in the SM way (solid curve in Figure~12b) and so
is already excluded by LEPI.  Additional radiative corrections due to
non-zero $A_t$ values and/or larger top and stop masses could increase
the lightest scalar mass beyond current limits. Such additional
radiative corrections are discussed in section 6.3.

We can take the MSSM limit of our results in a much more direct way  by
removing the $N$ components of the Higgs mass matrices by hand, and
numerically taking $\lambda$ and $k$ to be very small, and $r$ to be
very large, with $\mu$ selected arbitrarily and appearing in the
radiative corrections to the $2\times 2$ mass matrices instead of
$\lambda x$. The resulting spectrum in Figure~13a, for $\mu=0$ and
$\tan\beta=1.7$  is quite similar to that in Figure~12a, once the
lightest pseudoscalar and second lightest scalar of that plot have been
removed. The range of $m_c$ is extended with $m_c>80$ GeV as expected.
The lightest scalar again has a mass $m_h\approx 50$GeV with
$R_{ZZh}=\sin(\beta-\alpha) \approx 1$.

In Figure~13b we show the corresponding MSSM plots with $\tan\beta =10$,
and all other parameters as before. The limit of $\tan\beta\to\infty$ is
of course where the MSSM bound is saturated. In this limit the $2\times
2$ scalar mass-squared matrix is diagonal with elements $m_A^2$ and
$M_Z^2$ (plus radiative corrections). Furthermore,
$m_c^2=m_A^2+M_W^2$ at tree-level. This explains why the spectrum
in Figure~13b consists of a flat scalar mass plus a rising scalar mass
degenerate with the pseudoscalar. In this case, since $\beta\approx
\pi/2$ and $\alpha\approx 0$ (as the scalar matrix is diagonal), it is
clear that $R_{ZZh}=sin(\beta-\alpha)\approx 1$  and $R_{ZAh}=cos(\beta
-\alpha) \approx 0$,  where $h$ is defined to be the flat curve in
Figure~13b ({\it id est,} for values of $m_c$ less than the point at which
the two scalar curves bounce off each other, the lighter of the two
scalars is decoupled, while for larger values of $m_c$ the converse is
true.)

\smallheading{6.3 The Effect of Radiative Corrections}
All the spectra plotted in section 6.2 include top quark and squark
radiative
corrections as well as Higgs and Higgsino radiative corrections. However
in all these plots $A_t=0$, with $m_t=m_{\tilde t_1}=150GeV, m_{\tilde
t_2}=500GeV$. We shall now examine the effect of stop squark spectra
with $A_t\neq 0$. To understand how important the effect of non-zero
$A_t$ may be, we shall begin by considering the effects of each of the
other radiative corrections in turn. Since these effects are far harder
to compare accurately on a logarithmic scale, we shall restrict our
discussion to the lightest Higgs boson masses which we shall plot on a
linear scale. We shall use the  baseline parameters defined in \baseline,
and for comparison we shall in each case plot the
tree-level masses as dotted lines, so that the effect of the particular
radiative correction is easily seen.

Figure~14 shows the lightest Higgs boson masses corresponding to the
baseline parameters above, so that the Higgs masses are identical to
those in Figure~6a. The corresponding tree-level masses without top,
stop, Higgs or Higgsino radiative corrections are indicated by dotted
lines.  The lightest scalar mass is unaffected by radiative
corrections at its peak, where it decouples, but at this point the
corrections are largest for the next-to-lightest scalar. Away from
this point the corrections are shared between the two scalars, with
the pseudoscalar -- which is only weakly coupled across the whole
range -- receiving small corrections. Radiative corrections in
this case increase the second lightest CP-even scalar mass by as much
as 7 GeV.

Figure~15 shows the effect of radiative corrections due to top quark
loops alone, corresponding to the parameters and notation of Figure~14,
but with the Higgs and Higgsino radiative corrections subtracted. It is
clear by comparing Figure~15 to Figure~14 that the Higgs and Higgsino
corrections have relatively little effect compared to top quark loops,
the most noticeable change being felt by the pseudoscalar mass. As we
shall see, the Higgs and Higgsino radiative corrections are individually
quite large but tend to cancel between themselves (the cancellation
being exact in the limit of exact supersymmetry). Top quark loops are
therefore responsible for the bulk of the radiative corrections for the
baseline parameters.

Figure~16 shows the effect of radiative corrections due to Higgs
bosons alone, corresponding to the parameters and notation of
Figure~14, but this time with only the Higgs boson radiative
corrections included.  Compared to the tree-level results (dots), the
Higgs radiative corrections can be quite substantial and tend to
depress the scalar masses and enhance the pseudoscalar mass. Figure~17
shows the effect of radiative corrections due to Higgsinos alone for
the usual parameters and notation. In Figure~17 the comparison to the
tree-level dotted lines shows that Higgsinos tend to enhance scalar
masses in such a way as to partially cancel the effects of Higgs
corrections in Figure~16. The pseudoscalar mass is also enhanced in
this case, however.

We are now ready to examine the radiative corrections due to squark
spectra with $A_t\neq 0$. Again we use the baseline parameters in
\baseline, apart from the following two sets of stop parameters which
were encountered previously in our discussion of the bound in section
5:

(a) $m_{\tilde t_1}= 150 \GeV, m_{\tilde t_2}=500 \GeV , A_t= 700 \GeV$

(b) $m_{\tilde t_1}=600 \GeV, m_{\tilde t_2}=1 \TeV ,A_t= 1.8 \TeV$

Parameter set (a) uses the same stop masses as in the baseline parameters
but now involves a non-zero $A_t$ value chosen to maximise the effect of
radiative corrections. Parameter set (b) involves heavier stop masses
and larger $A_t$ value,  again chosen to maximise the effects of
radiative corrections. In Figure~18a we show the Higgs spectrum for
parameter set (a), including the radiative corrections due to the squark
spectrum as well as top, Higgs and Higgsino loops, with the
corresponding tree-level  spectrum again represented by dotted lines.
We see that the effect of the
$A_t=700$ GeV value in Figure~18a is dramatic for the physical second
lightest scalar with radiative corrections to its mass of up to 18 GeV
as compared to 7 GeV with $A_t=0$ in Figure~14 (with all other
parameters the same in the two cases).

In Figure~18b we show the spectrum for the heavier squark spectrum set
(b) above, again including all other radiative corrections in this plot
and comparing it to the baseline tree-level masses represented by dots.
The physical second lightest scalar in Figure~18b has radiative
corrections  to its mass of up to 25 GeV  as compared to 18 GeV in
Figure~18a. We conclude that the effect of choosing larger stop masses
and large non-zero $A_t$ values may significantly raise the mass of the
lightest physical Higgs bosons from their values shown in the plots of
section 6.2.

\num=0
\sec=7
\heading{7 Conclusion}
We have performed a comprehensive analysis of the radiative corrections
to the Higgs boson mass spectrum in the NMSSM using the one-loop
effective potential. Analytic results for top quark and squark
corrections are reviewed, and our numerical procedure for including
Higgs and Higgsino corrections is described. The bound on the lightest
CP-even scalar, including the radiative corrections mentioned above, was
then discussed. In order to find the absolute bound as a function of
$m_t$ including all radiative corrections, we maximised over the
parameter space using analytic and numerical techniques. Our final
results are summarised in Figure~4.

We have presented a numerical analysis of the RG flows for the
dimensionless couplings, relying on the assumption of perturbative
behaviour up to $M_{GUT}$. We emphasise the infra-red fixed points of
the gaugeless limit are not the relevant points for the low energy
behaviour because they are washed away by the large QCD corrections,
and instead the parameter space is limited to a region of the
$(\lambda/h_t)-(k/h_t)$ plane bounded by the axes and an approximate
quarter circle, as shown in Figure~1b.

A general discussion of Higgs boson phenomenology including radiative
corrections was then given. Since the parameter space of the Higgs
sector of the NMSSM is multi-dimensional, we selected a baseline set
of parameters given in \baseline, and discussed the effect of varying
each of the parameters in turn. Our selection of $k$ and $\lambda$ was
based on our study of RG flows described above. We found that larger
$m_c$ values are associated with larger values of $\tan\beta$ and $r$.
The NMSSM with small $k$ and $\lambda$ and large $r$ was also compared
to the MSSM (see Figures~12 and 13).  Possible characteristic
signatures of the NMSSM include light charged scalars, and weakly
coupled light neutral scalars, and there are regions of parameter
space where either or both signals is possible (see Figures~7 and 9).
Equally there are other regions of parameter space where neither of
these characteristic signals is present (see Figures~8 and 10).

Finally Figures~14-18 show the relative sizes of the various radiative
corrections. These show that the dominant effect is usually that of the
top quark in conjunction with the stop sector. Figure~18 emphasises the
huge radiative corrections which are possible for large $A_t$ and
non-degenerate stop squarks. Higgs and Higgsino corrections are
typically rather smaller, and tend to cancel due to the (softly broken)
supersymmetry, as shown in Figures~16 and 17. This cancellation also
occurs in the top-stop sector, but is less noticeable due to the large
soft squark masses which are assumed.

If Higgs bosons are discovered at LEP or LHC/SSC, then an important
question will be whether they are associated with the MSM, the MSSM, or
some other model. If they are not associated with the MSSM, then it is
possible that they arise from the NMSSM, which has a more general
structure. The discussion of radiative corrections to Higgs boson masses
in the NMSSM and the phenomenological discussion presented in this
paper will help to decide this question.

\heading{Acknowledgements}
We are grateful to the SERC for financial support. PLW would like to
thank Steve Kelley for some very helpful suggestions about effective
potential techniques.

\vfill
\eject

\heading{References}
\parindent -15pt

[\ref{\reviews}] H.P.~Nilles, {\it Phys. Rep.}
                {\bf 110} (1984) 1; \hfil\break
\quad H.E.~Haber and G.L.~Kane, {\it Phys. Rep.} {\bf 117} (1985) 75.

[\ref{\hhguide}] J.F.~Gunion, H.E.~Haber, G.L.~Kane and S.~Dawson,
  ``The Higgs Hunter's Guide'' (Addison-Wesley, Reading, MA, 1990).

[\ref{\okada}]  H.~Haber and R.~Hempfling, {\it Phys. Rev. Lett.}
     {\bf 66} (1991) 1815; \hfil\break
\qquad Y.~Okada, M.~Yamaguchi and T.~Yanagida, {\it Prog. Theor. Phys.}
   {\bf 85} (1991) 1; {\it Phys. Lett.} {\bf B262} (1991) 54.

[\ref{\erz}] J.~Ellis, G.~Ridolfi and F.~Zwirner, {\it Phys. Lett.}
   {\bf B257} (1991) 83; {\it Phys. Lett.} {\bf B262} (1991) 477;
\hfil\break
\qquad A.~Brignole, J.~Ellis, G.~Ridolfi and F.~Zwirner, {\it Phys. Lett.}
   {\bf B271} (1991) 123.

[\ref{\brignole}] A.~Brignole, {\it Phys. Lett.} {\bf B277} (1992) 313;
     {\it Phys. Lett.} {\bf B281} (1992) 284.

[\ref{\eahd}] J.L.~Lopez and D.V.~Nanopoulos, {\it Phys. Lett.}
         {\bf B266} (1991) 397; \hfil\break
\qquad K.~Sasaki, M.~Carena and C.E.M.~Wagner, {\it Nucl. Phys.}
               {\bf B381} (1992) 66.

[\ref{\haber}] H.E.~Haber and R.~Hempfling, preprint number SCIPP 91/33.

[\ref{\fayet}] P.~Fayet, {\it Nucl. Phys.} {\bf B90} (1975) 104.

[\ref{\NMSSM}] J.~Ellis, J.F.~Gunion, H.E.~Haber, L.~Roszkowski and
   F.~Zwirner, {\it Phys. Rev.} {\bf D39} (1989) 844.

[\ref{\dl}] L.~Durand and J.L.~Lopez, {\it Phys. Lett.}
       {\bf B217} (1989) 463; \hfil\break
\qquad L.~Drees, {\it Int. J. Mod. Phys.} {\bf A4} (1989) 3635.

[\ref{\eq}] J.R.~Espinosa and M.~Quiros,
     {\it Phys. Lett.} {\bf B279} (1992) 92.

[\ref{\kkw}] G.~Kane, C.~Kolda and J.~Wells, {\it Phys. Rev. Lett.}
       {\bf 70} (1993) 2686.

[\ref{\eltraub}] U.~Ellwanger and M.~Rausch de Traubenberg, {\it Z. Phys.}
        {\bf C53} (1992) 521.

[\ref{\ellind}] U.~Ellwanger and M.~Lindner, {\it Phys. Lett.} {\bf B301}
         (1993) 365.

[\ref{\el}] U.~Ellwanger, {\it Phys. Lett.} {\bf B303} (1993) 271.

[\ref{\terV}] W.~ter Veldhuis, Purdue preprint PURD-TH-92-11 and
     hep-ph/9211281.

[\ref{\eqtwo}] J.R.~Espinosa and M.~Quiros, {\it Phys. Lett.} {\bf B302}
    (1993) 51.

[\ref{\ekwone}] T.~Elliott, S.F.~King and P.L.~White, {\it Phys. Lett.}
 {\bf B305} (1993) 71.

[\ref{\ekwtwo}] T.~Elliott, S.F.~King and P.L.~White, Southampton preprint
 SHEP 92/93-18, {\it Phys. Lett.} to be published.

[\ref{\pandita}] P.N.~Pandita, Northeastern Hill University preprint
       PRINT-93-0465, {\it Z.Phys.} to be published.

[\ref{\cw}] S.~Coleman and E.~Weinberg, {\it Phys. Rev.} {\bf D7} (1973) 1888;
      \hfil \break
\qquad S.~Weinberg, {\it Phys. Lett.} {\bf D7} (1973) 2887.

[\ref{\sher}] M.~Sher, {\it Phys. Rep.} {\bf 179} (1989) 273.

[\ref{\stabone}] S.~Ferrara, D.V.~Nanopoulos and C.A.~Savoy,
          {\it Phys. Lett.} {\bf B123} (1983) 214; \hfil\break
\qquad J.~Polchinski and L.~Susskind, {\it Phys. Rev.} {\bf 26} (1982) 3661;
                 \hfil\break
\qquad H.P.~Nilles, M.~Srednicki and D.~Wyler, {\it Phys. Lett.} {\bf B124}
      (1983) 337; \hfil\break
\qquad A.B.~Lahanes, {\it Phys. Lett.} {\bf B124}
      (1983) 341; \hfil\break
\qquad L.~Alvarez-Gaume, J.~Polchinski and M.B.~Wise, {\it Nucl. Phys.}
      {\bf B221} (1983) 495.

[\ref{\stabtwo}] J.~Bagger and E.~Poppitz, John Hopkins preprint number
 JHU-TIPAC-93018 and hep-ph 9307317.

[\ref{\ds}] J.-P.~Derendinger and C.A.~Savoy, {\it Nucl. Phys.} {\bf B237}
   (1984) 307.

[\ref{\bs}] P.~Binetruy and C.A.~Savoy, {\it Phys. Lett.}
   {\bf B277} (1992) 453.

[\ref{\numrec}] S.~Jacoby, J.~Kowalik and J.~Pizzo, ``Iterative Methods for
   Non-Linear Optimization Problems'' (Prentiss-Hall, 1972); \hfil\break
\qquad W.H.~Press, B.P.~Flannery, S.A.~Teukolsky and W.~T.Vetterling,
 ``Numerical Recipes'' (Cambridge University Press, 1986).

[\ref{\roger}] V.~Barger, K.~Cheung, R.J.N.~Phillips and A.L.~Stange,
      {\it Phys. Rev.} {\bf D46} (1992) 4914; \hfil\break
\qquad H.~Baer {\it et al},
      {\it Phys. Rev.} {\bf D46} (1992) 1067; \hfil\break
\qquad J.~Gunion {\it et al},
      {\it Phys. Rev.} {\bf D46} (1992) 2040, 2052; \hfil\break
\qquad J.~Gunion {\it et al},
      {\it Phys. Rev.} {\bf D47} (1993) 1030; \hfil\break
\qquad Z.~Kunszt and F.~Zwirner,
      {\it Nucl. Phys.} {\bf B285} (1992) 3.

[\ref{\bsg}] J.L.~Hewett,
      {\it Phys. Rev. Lett.} {\bf 70} (1993) 1045; \hfil\break
\qquad V.~Barger, M.~Berger and R.J.N.~Phillips,
      {\it Phys. Rev. Lett.} {\bf 70} (1993) 1368.

[\ref{\barb}] R.~Barbieri and G.F.~Giudice, CERN preprint CERN-TH.6830/93.

[\ref{\eltwo}] U.~Ellwanger, M.~Rausch de Traubenberg and C.A.~Savoy,
     Heidelberg preprint HD-THEP-93-95 and hep-ph/9307322.

\vfill
\eject

\num=0
\sec=4
\eject
\heading{Figure Captions}

\noindent
{\bf Figure 1a: }
Renormalisation group flow of a set of points in the
$(\lambda/h_t)$--$(k/h_t)$ plane as the energy scale is varied from
$10^{16}$GeV to 1TeV. $h_t(10^{16}\hbox{GeV})=10$ and gauge
couplings have been set to zero.

\bigskip
\noindent
{\bf Figure 1b: }
Renormalisation group flow of a set of points in the
$(\lambda/h_t)$--$(k/h_t)$ plane as the energy scale is varied from
$10^{16}$GeV to 1TeV. $h_t(10^{16}\hbox{GeV})=10$ and gauge
couplings have been included.

\bigskip
\noindent
{\bf Figure 1c: }
Renormalisation group flow of a set of points in the
$(\lambda/h_t)$--$(k/h_t)$ plane as the energy scale is varied from
$10^{16}$GeV to 1TeV. $h_t(10^{16}\hbox{GeV})=1$ and gauge
couplings have been included.

\bigskip
\noindent
{\bf Figure 2: }
$\lambda_{\rm max}$ against $h_t(1{\rm TeV})$

\bigskip
\noindent
{\bf Figure 3: }
The values of $\sin\beta$ (long dashes), $\lambda$
(short dashes), and $h_t$ (solid line) for which the bound is
maximised, for the value $m_{\tilde t_2}=1$TeV. Results for other
values of $m_{\tilde t_2}$ are very similar.

\bigskip
\noindent
{\bf Figure 4: }
The bound on the lightest neutral CP-even scalar against $m_t$. The
two solid lines are for the NMSSM; the dotted lines for the MSSM. In
each case the upper (lower) line is for $m_{\tilde t_2}=$1TeV
(500GeV).

\bigskip
\noindent
{\bf Figure 5: }
The bound on the lightest scalar against $m_{\tilde
t_1}$ expressed in units of $m_{\tilde t_2}$. The upper (lower) line
represents the case where $m_{\tilde t_2}$=1TeV (500GeV).

\bigskip
\noindent
{\bf Figure 6a: }
Higgs boson masses against $m_c$ in the NMSSM.  The
parameters are $r=1.0$, $\tan\beta=1.7$, $\lambda=0.65$, $k=0.1$,
$A_k=0$, $m_t=150$GeV, $m_{\tilde t_1}= 150$ GeV, $m_{\tilde t_2}=500$
GeV, and $A_t=0$.  Solid lines represent CP-even, and dashed lines
CP-odd mass eigenstates.

\bigskip
\noindent
{\bf Figure 6b: }
Higgs boson couplings and N-amplitudes corresponding
to Fig.6a.  The solid line represents the $R_{ZZh}$ coupling, the
dashed line the $R_{ZhA}$ coupling. The amplitude of singlet field $N$
contained in each of the mass eigenstates are indicated by a dotted
line for the lightest scalar $h$, a dot-dashed line for the
next-to-lightest scalar, and a dot-dot-dashed line for the lightest
pseudo-scalar $A$.

\bigskip
\noindent
{\bf Figure 7a: }
Higgs boson masses against $m_c$ in the NMSSM.  The
parameters are as in Fig.6a except that $r=0.1$.  Solid lines
represent CP-even, and dashed lines CP-odd mass eigenstates.

\bigskip
\noindent
{\bf Figure 7b: }
Higgs boson couplings and N-amplitudes corresponding
to Fig.7a.  The solid line represents the $R_{ZZh}$ coupling, the
dashed line the $R_{ZhA}$ coupling. The amplitude of singlet field $N$
contained in each of the mass eigenstates are indicated by a dotted
line for the lightest scalar $h$, a dot-dashed line for the
next-to-lightest scalar, and a dot-dot-dashed line for the lightest
pseudo-scalar $A$.

\bigskip
\noindent
{\bf Figure 8a: }
Higgs boson masses against $m_c$ in the NMSSM.  The
parameters are as in Fig.6a except that $r=5.0$.  Solid lines
represent CP-even, and dashed lines CP-odd mass eigenstates.

\bigskip
\noindent
{\bf Figure 8b: }
Higgs boson couplings and N-amplitudes corresponding
to Fig.8a.  The solid line represents the $R_{ZZh}$ coupling, the
dashed line the $R_{ZhA}$ coupling. The amplitude of singlet field $N$
contained in each of the mass eigenstates are indicated by a dotted
line for the lightest scalar $h$, a dot-dashed line for the
next-to-lightest scalar, and a dot-dot-dashed line for the lightest
pseudo-scalar $A$.

\bigskip
\noindent
{\bf Figure 9a: }
Higgs boson masses against $m_c$ in the NMSSM.  The
parameters are as in Fig.6a except that $\tan\beta=10$.  Solid lines
represent CP-even, and dashed lines CP-odd mass eigenstates.

\bigskip
\noindent
{\bf Figure 9b: }
Higgs boson couplings and N-amplitudes corresponding
to Fig.9a.  The solid line represents the $R_{ZZh}$ coupling, the
dashed line the $R_{ZhA}$ coupling. The amplitude of singlet field $N$
contained in each of the mass eigenstates are indicated by a dotted
line for the lightest scalar $h$, a dot-dashed line for the
next-to-lightest scalar, and a dot-dot-dashed line for the lightest
pseudo-scalar $A$.

\bigskip
\noindent
{\bf Figure 10a: }
Higgs boson masses against $m_c$ in the NMSSM.
The parameters are as in Fig.6a except that $\lambda=0.4$, $k=0.4$.
Solid lines represent CP-even, and dashed lines CP-odd
mass eigenstates.

\bigskip
\noindent
{\bf Figure 10b: }
Higgs boson couplings and N-amplitudes
corresponding to Fig.10a.  The solid line represents the $R_{ZZh}$
coupling, the dashed line the $R_{ZhA}$ coupling. The amplitude of
singlet field $N$ contained in each of the mass eigenstates are
indicated by a dotted line for the lightest scalar $h$, a dot-dashed
line for the next-to-lightest scalar, and a dot-dot-dashed line for
the lightest pseudo-scalar $A$.

\bigskip
\noindent
{\bf Figure 11a: }
Higgs boson masses against $m_c$ in the NMSSM.  The
parameters are as in Fig.6a except that $\lambda=0.1$, $k=0.6$.  Solid
lines represent CP-even, and dashed lines CP-odd mass eigenstates.

\bigskip
\noindent
{\bf Figure 11b:}
Higgs boson couplings and N-amplitudes corresponding
to Fig.11a.  The solid line represents the $R_{ZZh}$ coupling, the
dashed line the $R_{ZhA}$ coupling. The amplitude of singlet field $N$
contained in each of the mass eigenstates are indicated by a dotted
line for the lightest scalar $h$, a dot-dashed line for the
next-to-lightest scalar, and a dot-dot-dashed line for the lightest
pseudo-scalar $A$.

\bigskip
\noindent
{\bf Figure 12a: }
Higgs boson masses against $m_c$ in the NMSSM.  The
parameters are as in Fig.6a except that $\lambda=0.1$, $k=0.1$.  Solid
lines represent CP-even, and dashed lines CP-odd mass eigenstates.

\bigskip
\noindent
{\bf Figure 12b: }
Higgs boson couplings and N-amplitudes
corresponding to Fig.12a.  The solid line represents the $R_{ZZh}$
coupling, the dashed line the $R_{ZhA}$ coupling. The amplitude of
singlet field $N$ contained in each of the mass eigenstates are
indicated by a dotted line for the lightest scalar $h$, a dot-dashed
line for the next-to-lightest scalar, and a dot-dot-dashed line for
the lightest pseudo-scalar $A$.

\bigskip
\noindent
{\bf Figure 13a: }
Higgs boson masses against $m_c$ for the MSSM.  The
parameters are $\tan\beta=1.7$, $m_t=150$GeV, $m_{\tilde t_1}= 150$
GeV, $m_{\tilde t_2}=500$ GeV, $\mu=0$ and $A_t=0$.  CP-even and
CP-odd mass eigenstates are represented by solid and dashed lines
respectively.

\bigskip
\noindent
{\bf Figure 13b: }
Higgs boson masses against $m_c$ for the MSSM.  The
parameters are as in Fig.13a except that $\tan\beta=10$.  CP-even and
CP-odd mass eigenstates are represented by solid and dashed lines
respectively.

\bigskip
\noindent
{\bf Figure 14: }
Higgs boson masses against $m_c$ in the NMSSM.  The
parameters are $r=1.0$, $\tan\beta=1.7$, $\lambda=0.65$, $k=0.1$,
$A_k=0$, $m_t=150$GeV, $m_{\tilde t_1}= 150$ GeV, $m_{\tilde t_2}=500$
GeV, and $A_t=0$.  Solid lines represent CP-even, and dashed lines
CP-odd mass eigenstates and include radiative corrections due to loops
of top quarks, squarks, Higgs bosons and Higgsinos.  This plot is as
in Fig.6a except that it is rescaled to show the lighter mass
eigenstates in greater detail.  The dotted lines indicate the
tree-level spectrum for the same parameters.

\bigskip
\noindent
{\bf Figure 15: }
Higgs boson masses against $m_c$ in the NMSSM.  The
parameters are as in Fig.14.  Solid lines represent CP-even, and
dashed lines CP-odd mass eigenstates and include radiative corrections
due to loops of top quarks and squarks only.  The dotted lines
indicate the tree-level spectrum for the same parameters.

\bigskip
\noindent
{\bf Figure 16: }
Higgs boson masses against $m_c$ in the NMSSM.  The
parameters are as in Fig.14.  Solid lines represent CP-even, and
dashed lines CP-odd mass eigenstates and include radiative corrections
due to loops of Higgs bosons only.  The dotted lines indicate the
tree-level spectrum for the same parameters.

\bigskip
\noindent
{\bf Figure 17: }
Higgs boson masses against $m_c$ in the NMSSM.  The
parameters are as in Fig.14.  Solid lines represent CP-even, and
dashed lines CP-odd mass eigenstates and include radiative corrections
due to loops of Higgsinos only.  The dotted lines indicate the
tree-level spectrum for the same parameters.

\bigskip
\noindent
{\bf Figure 18a: }
Higgs boson masses against $m_c$ in the NMSSM.  The
parameters are as in Fig.14 except that $A_t=700GeV$.  Solid lines
represent CP-even, and dashed lines CP-odd mass eigenstates and
include radiative corrections due to loops of top quarks, squarks,
Higgs bosons and Higgsinos.  The dotted lines indicate the tree-level
spectrum for the same parameters.

\bigskip
\noindent
{\bf Figure 18b: }
Higgs boson masses against $m_c$ in the NMSSM.  The
parameters are as in Fig.14 except that $m_{\tilde t_1}=600$ GeV,
$m_{\tilde t_2}=1.0$ TeV, $A_t=1.8$ TeV.  Solid lines represent
CP-even, and dashed lines CP-odd mass eigenstates and include
radiative corrections due to loops of top quarks, squarks, Higgs
bosons and Higgsinos.  The dotted lines indicate the tree-level
spectrum for the same parameters.

\vfill
\eject

\bye